\documentclass[useAMS,usenatbib]{mn2e}
\usepackage{graphicx}
\def\lesssim{\!\!\!\phantom{\le}\smash{\buildrel{}\over
 {\lower2.5dd\hbox{$\buildrel{\lower2dd\hbox{$\displaystyle<$}}\over
                               \sim$}}}\,\,}
\title[SN 2006jc: new and old dust in the progenitor CSM.]
{Massive stars exploding in a He-rich circumstellar medium III. SN 2006jc: 
IR~echoes from new and old dust in the progenitor CSM.}
\author[Mattila et al.]
{S. Mattila\thanks{s.mattila@utu.fi}$^{1,2}$,W.P.S. Meikle$^{3}$,
P. Lundqvist$^{4}$, A. Pastorello$^{1}$, R. Kotak$^{1}$, J. Eldridge$^{5}$,\and
S. Smartt$^{1}$, A. Adamson$^{6}$, C.L. Gerardy$^{7}$, L. Rizzi$^{6}$, 
A. W. Stephens$^{8}$, S.D. Van Dyk$^{9}$\\
$^{1}$ Astrophysics Research Centre, School of Mathematics and Physics, 
Queen's University Belfast, BT7 1NN, UK.\\ 
$^{2}$ Tuorla Observatory, University of Turku, V\"ais\"al\"antie 20, FI-21500 
Piikki\"o, Finland.\\ 
$^{3}$ Astrophysics Group, Blackett Laboratory, Imperial College London, Prince 
Consort Road, London SW7 2AZ, UK.\\ 
$^{4}$ Department of Astronomy, Stockholm University, AlbaNova, SE-106 91 
Stockholm, Sweden.\\ 
$^{5}$ Institute of Astronomy, The Observatories, University of Cambridge,
Madingley Road, Cambridge, CB3 0HA.\\
$^{6}$ Joint Astronomy Centre, 660 North A'ohoku Place, Hilo, HI 96720, USA.\\
$^{7}$ Department of Physics, Florida State University, Tallahassee, FL 32306,
USA.\\
$^{8}$ Gemini Observatory, 670 North A'ohoku Place, Hilo, HI 96720.\\
$^{9}$ Spitzer Science Center/Caltech, Pasadena, CA 91125, USA.\\}

\date{Accepted ???? December ??.
      Received ???? December ??;
      in original form ???? October ??}
\begin{document}

\maketitle

\label{firstpage}

\begin{abstract}
We present near- and mid-infrared (IR) photometric data of the Type
Ibn supernova (SN) 2006jc obtained with the United Kingdom Infrared
Telescope (UKIRT), the Gemini North Telescope, and the Spitzer Space
Telescope between days 86 and 493 post-explosion.  We find that the IR
behaviour of SN~2006jc can be explained as a combination of IR~echoes
from two manifestations of circumstellar material.  The bulk of the
near-IR emission arises from an IR~echo from newly-condensed dust in a
cool dense shell (CDS) produced by the interaction of the ejecta
outward shock with a dense shell of circumstellar material ejected by
the progenitor in a luminous blue variable (LBV) like outburst about
two years prior to the SN explosion.  The CDS dust mass reaches a
modest 3.0$\times$10$^{-4}$ M$_{\odot}$ by day~230.  While dust
condensation within a CDS formed behind the ejecta inward shock has
been proposed before for one event (SN~1998S), SN 2006jc is the first
one showing evidence for dust condensation in a CDS formed behind the
ejecta outward shock in the circumstellar material. At later epochs, a
substantial and growing contribution to the IR fluxes arises from an
IR~echo from pre-existing dust in the progenitor wind. The mass of the
pre-existing CSM dust is at least
$\sim8\times$10$^{-3}$~M$_{\odot}$. This work therefore adds to
the evidence that mass-loss from the {\it progenitors} of
core-collapse supernovae could be a major source of dust in the
universe. However, yet again, we see no direct evidence that the
explosion of a supernova produces anything other than a very modest
amount of dust.
\end{abstract}


\section{Introduction}
The importance of core-collapse supernovae (CCSNe) as a source of
cosmic dust is currently a highly debated topic. For several decades
it has been hypothesized that the physical conditions in the expanding
SN ejecta could result in the condensation of large amounts of dust
(e.g., Cernuschi, Marsicano, \& Codina 1967; Hoyle \& Wickramasinghe
1970; Gehrz 1989; Tielens 1990; Dwek 1998).  More recently, CCSNe
arising from Population III stars have been proposed as the main
source of dust in the early universe (Todini \& Ferrara 2001; Nozawa
et al. 2003, Dwek et al. 2007). Models of dust formation in CCSNe
(Todini \& Ferrara 2001; Nozawa et al. 2003, 2008) succeed in
producing large amounts of dust that would be enough to account for
the dust seen at high redshifts (see Meikle et al. 2007). However,
direct observational evidence for CCSNe as a major source of dust is
still missing, even in the local universe (Meikle et al. 2007 and
references therein).\\

SN~2006jc was discovered on 2006 October 9.75 UT by Nakano et
al. (2006) in the nearby spiral galaxy UGC~4904
and was classified as a peculiar Type Ib SN (Crotts et
al. 2006; Benetti et al. 2006; Modjaz et al. 2006).  The supernova was
discovered after optical maximum.  However, model fits to
the bolometric light curve (Pastorello et al.  2008a) yielded the most
satisfactory fits with an explosion date of 2006 September 21
(JD=2454000). In the following paper we shall adopt this date as epoch
t=0.  Comparison with the earlier discovered SN 1999cq suggests that
optical maximum occurred at about +8 to +10~days (Pastorello et
al. 2008a).  The early-time SN shows an apparently hybrid spectrum
with broad emission lines of intermediate mass elements commonly
observed in Type Ic SNe and relatively narrow (FWHM $\sim$ 2000--3000
km/s) emission lines of helium originating from a dense CSM around the
SN (Foley et al. 2007; Pastorello et al. 2007). The He~I lines were
already apparent in the first spectrum obtained at $\sim$20 days and
persisted until at least 180 days (Pastorello et al. 2008a). In
addition, SN 2006jc showed H$\alpha$ emission with a narrower profile
indicating an origin in a different CSM region from that which gave
rise to the He lines. Excess emission in both UV and X-rays (Brown et
al. 2006; Immler et al. 2006, 2008) also indicates the presence of a
substantial CSM. It appears that SN~2006jc actually belongs to a
sub-class of Type~Ic events which show evidence of a dense He-rich
CSM. Other examples are SNe~1999cq and 2002ao (Matheson et al. 2000;
Foley et al. 2007; Pastorello et al 2008a), SN~2000er (Pastorello et al. 
2008a) and SN~2005la which appears to be a transitional case between SN
2006jc-like events and Type IIn SNe (Pastorello et al. 2008b).  A new
classification as Type Ibn has been proposed (Pastorello et al. 2007;
Pastorello et al. 2008a) for such SN 2006jc-like events.\\

An outburst similar to those exhibited by the most energetic Luminous
Blue Variables (LBV) was detected at the SN 2006jc location two years
before its explosion (Nakano et al. 2006; Pastorello et
al. 2007). Foley et al. (2007) and Pastorello et al. (2007) suggested
that a helium-rich shell was ejected during this event and that this shell 
is giving rise to the He~I lines. The apparent LBV-like outburst indicates 
that the progenitor of both the outburst and SN 2006jc might have been a very 
massive star (Foley et al. 2007; Pastorello et al. 2007; Pastorello et al.
2008a). Alternatively, SN 2006jc could have originated in a binary 
system consisting of an LBV that erupted in 2004, and a Wolf-Rayet star that 
gave rise to SN 2006jc (Pastorello et al. 2007, 2008a).\\

SN 2006jc has provided also the first ever opportunity of
observing dust formation associated with this subtype of CCSN.  Dust
production associated with SNe can be studied via the thermal infrared
(IR) emission from the grains, or by their attenuating effect on light
passing through the dusty regions. Near-IR (NIR) excesses have been
observed in five Type IIn SNe and five other Type~II subtypes (e.g.,
Fassia et al. 2000; Di Carlo et al. 2002; Gerardy et
al. 2002). However, prior to SN~2006jc, only in three examples of
non-Type~II core-collapse SNe have NIR excesses been reported: SN
1982E (probable Type~Ib, Graham \& Meikle 1986), SN~1982R (Type Ib,
Graham 1985; Graham \& Meikle 1986) and SN~2002ic (peculiar event,
Kotak et al. 2004).  The attenuation method has been applied to the
Type~IIpec SN~1987A (e.g., Danziger et al. 1989; Lucy et al. 1989), 
the Type~Ib SN 1990I (Elmhamdi et al. 2004), the Type~IIn SN~1998S 
(Pozzo et al. 2004), and the Type~IIP SNe~1999em (Elmhamdi et al. 2003) 
and 2003gd (Sugerman et al. 2006).\\

As early as +55 days, SN 2006jc had already developed a strong NIR
excess (Arkharov et al. 2006; Minezaki et al. 2007; Smith et
al. 2008).  Observations by Di Carlo et al. (2008) and by us (see
below) show that the NIR excess peaked at around 80~days and persisted
to past 200~days.  Sakon et al.  (2008) report NIR and mid-IR (MIR)
observations at 220~days, confirming the persistence of the IR excess
to at least this epoch.  We confirm this, and find that the IR excess
persisted to at least 493~days.  In addition, optical
observations reported by Smith et al. (2008) and by us show that the
narrow He I lines became systematically blueshifted after $\sim$50
days and that over the same period an abrupt steepening was observed
in the optical light curves. The optical light curves of SN 2006jc are
also analysed by Tominaga et al. (2008), Di Carlo et al.  (2008) and
Pastorello et al. (2008a).\\

A study of the IR excess in SN~2006jc was first presented by
Smith et al. (2008).  Subsequent papers discussing the IR excess
include those of Sakon et al. (2008), Tominaga et al. (2008), Di Carlo
et al. (2008), Nozawa et al. (2008) and the present work. Smith et
al., Di Carlo et al. and the work presented here, all propose dust
formation in an outward shock-formed cool dense shell (CDS) to account
for the NIR emission.  In contrast, Sakon et al., Tominaga et al., and
Nozawa et al. propose dust formation in the SN ejecta.  The idea of
dust formation in a CDS in the CCSN context was originally introduced
by Pozzo et al. (2004) to account for the IR and optical behaviour of
SN~1998S. SN~2006jc provides the second opportunity to study this phenomenon.
Therefore, to investigate the origin of the IR excess in SN 2006jc, we commenced a
NIR and MIR photometric monitoring campaign via Director's
Discretionary Time (DDT) on the United Kingdom Infrared Telescope
(UKIRT) and the Gemini North Telescope, and Target of Opportunity
(ToO) observations with the Spitzer Space Telescope (Spitzer).\\

In this paper, we examine the presence of newly-formed dust in SN~2006jc via
both its IR emission and its attenuating effects on the optical
emission.  Using a more extensive IR dataset than presented in
previous studies, plus modelling of the shock interaction, we confirm
the proposition of Smith et al. and Di Carlo et al. of the dust
formation in a CDS and strengthen the support for it. In addition, we
show (a) how absorption and reradiation by the CDS dust of the
early-time UV/optical emission from the SN (i.e. an IR-echo) can
provide a self-consistent explanation for the bulk of the NIR energy
and evolution, and (b) show that a second, cooler IR echo also
occurred due to dust in the undisturbed progenitor CSM. Optical
observations and a systematic study of the observed properties of SN 2006jc 
and the four other Type Ibn events are presented in two companion papers 
(Pastorello et al. 2008a,b).

\begin{figure*}
\begin{minipage}{170mm}
\begin{center}
\includegraphics[width=150mm, angle=0, clip]{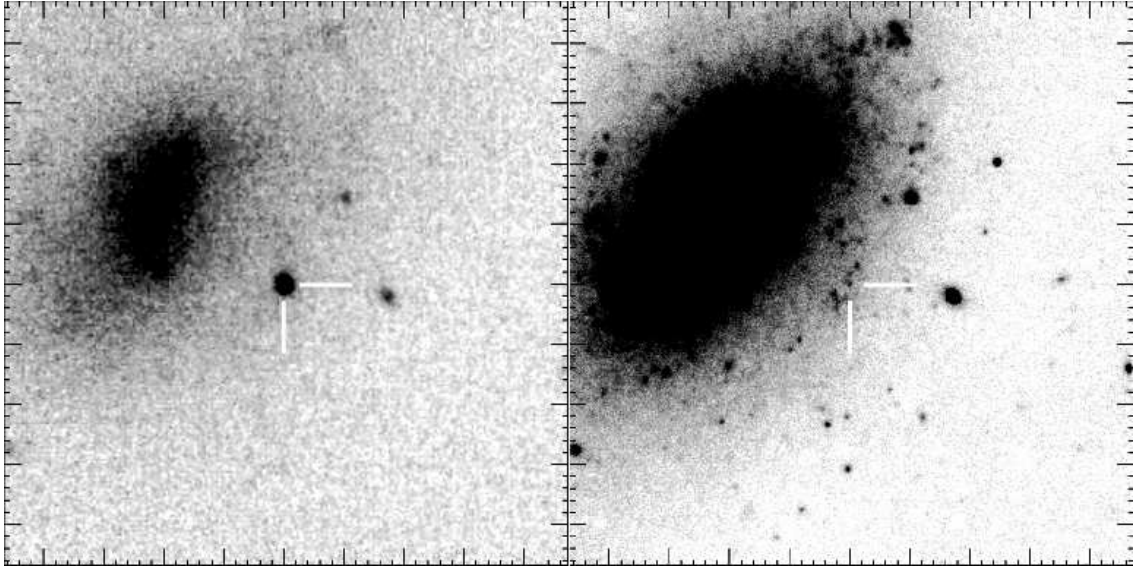}
\end{center}
\caption{The field of SN 2006jc at 2.2~$\mu$m. The FOV of the
image is 47 $\times$ 47 arcsec and the SN (marked with ticks)
is located in the middle. The left-hand image was obtained by combining 
UKIRT images from 2007 April 26 and May 10. The right-hand image is a 
deep integration obtained at Gemini on 2008 January 27. North is up and 
East to the left.}
\end{minipage}
\end{figure*}

\section{Infrared Observations}
SN 2006jc was observed in the $JHK$ bands with the WFCAM wide field NIR
imager on UKIRT at eight epochs between 2006 December 16 (epoch +86 days)
and 2007 May 10 (epoch +231 days). At each epoch, 5 dithered images were 
acquired with the SN placed both in the NW and SE corners of one of the 
WFCAM detectors. The data were reduced and photometrically calibrated via 
the WFCAM pipeline at the Cambridge Astronomical Survey Unit (CASU). A 
$K$-band observation was also obtained with the UFTI NIR imager on UKIRT 
on 2007 June 27 (epoch +279 days).\\

Seven months later, on 2008 January 27 (epoch +493 days), a deep 3840
sec on-source integration was obtained with the NIRI NIR imager on the
Gemini North Telescope under the DDT program GN-2007B-DD-8. The
jittered on-source frames were sky subtracted using the IRAF XDIMSUM
package and were median-combined, excluding a few frames with less
successful sky subtraction.  The final combined image has a seeing
FWHM of $\sim$0.45 arcsec. The photometric calibration utilised a
nearby photometric standard (FS 127) observed immediately after the
SN. To identify the SN location we aligned a combined (2007 April 26
and May 10) UKIRT $K$-band image with the Gemini image.  For this
procedure, 15 point-like sources common to the two images were used to
obtain a general geometric transformation (with no distortion term),
yielding an rms of 0.03 and 0.02 arcsec in $x$ and $y$,
respectively. The aligned UKIRT image and the deep Gemini image are
shown in Fig. 1. A faint point source is present in the Gemini image
coincident with the SN position.\\

SN 2006jc was also observed with the Spitzer's Infrared Array Camera
(IRAC) at 3.6, 4.5, 5.8, and 8.0~$\mu$m on 2007 May 7 (epoch +228 days)
and November 25 (epoch +430 days) within Spitzer programs PID 30292
and 40619. The pre-explosion field of SN 2006jc was also
serendipitously observed within the Spitzer Infrared Nearby Galaxies
Survey (SINGS) (PID: 0159; Kennicutt et al. 2003) at 3.6 and 5.8$\mu$m
on 2004 April 30th.  We used the post-basic calibrated data (PBCD)
products provided by the S16.1.0 version of the Spitzer pipeline in
this study. The pre- and post-explosion (2007 May 7) 5.8$\mu$m IRAC images 
are shown in Fig. 2.\\

\section{Infrared photometry of SN~2006jc}
Aperture photometry was performed on the two sets of UKIRT $JHK$
images obtained at each epoch, using the Starlink package GAIA
(Draper, Gray \& Perry 2002). A 3.0 arcsec radius aperture was used
for all except the latest epochs, where a 2.0 arcsec radius aperture
was used to yield more accurate photometry when the SN was faint
relative to the host galaxy. The sky was measured within a 1.5-2.0
$\times$ radius annulus. Aperture correction in each image was carried
out via large-aperture photometry of three nearby bright stars
(2MASS J09173378+4153251, J09172752+4153381 and J09170785+4152504 in the $J$
and $H$ bands, and J09173181+4151543, J09172752+4153381 and
J09170557+4154505 in the $K$-band). The magnitudes of the stars were
compared with their average values (over all the epochs) to check and
if necessary adjust the photometric calibration produced by the CASU
pipeline. We also compared these magnitudes with those available from
2MASS and found that they agreed within 0.06, 0.02 and 0.02 magnitudes
in $J$, $H$, and $K$ bands, respectively. Finally, the average of the
two measurements at each epoch was adopted as the SN magnitude. The
statistical error in the SN photometry and the standard deviation of
the recalibrated field star magnitudes were added in quadrature to
yield the uncertainty in each measurement. The resulting SN
photometric measurements are listed in Table~1.\\

To measure the SN magnitude in the Gemini $K$-band image we performed
point-spread function (PSF) fitting using the SNOOPY\footnote{SNOOPY, 
originally presented in Patat (1996), has been implemented in IRAF 
by E. Cappellaro. The package is based on DAOPHOT, but optimised for 
SN magnitude measurements.} package based on IRAF's DAOPHOT.  For this 
measurement the SN position was fixed according to the centroid coordinates 
obtained from the aligned UKIRT image where the SN was still bright.  The 
photometric uncertainty was estimated via PSF-fitting to artificial sources 
placed close to the SN position after subtracting the PSF-fit at the SN
position.  This yielded a $K$-band magnitude of 21.64 $\pm$ 0.40 for
the SN.\\

GAIA was also used to perform aperture photometry on the SN in the
Spitzer IRAC images. A 2.25 arcsec radius aperture was used and the
sky was measured within a 1.5-2.0 $\times$ radius annulus. This
aperture was chosen as a compromise between maximising the sampled
fraction of the source flux (the radius of the first diffraction minimum
at the extreme red end of the 8.0$\mu$m channel is 2.6'') and
minimising any extended residual emission in the subtracted images
(see below). Aperture corrections were derived from the IRAC PSF
images available on the Spitzer website.  The correction factors were
1.23, 1.26, 1.50, and 1.65 for 3.6, 4.5, 5.8, and 8.0$\mu$m,
respectively.  The contribution of background flux to these results
was assessed through the use of image subtraction.  At 3.6 and
5.8$\mu$m, we subtracted the pre-explosion SINGS images from our
post-explosion PBCD-processed images using image matching and
subtraction techniques as implemented in the ISIS 2.2
image-subtraction package (Alard 2000).  In Meikle et al. (2006), we
demonstrated the applicability of the image-subtraction technique for
Spitzer/IRAC SN data and assessed its uncertainties.  For SN~2006jc,
we found that for day~+228 the fluxes measured in the subtracted
frames were about 5\% lower than in the unsubtracted images.  However,
no point source was apparent in the IRAC pre-explosion images at the
SN location (see Fig. 2), such as might have been produced by the
presumably dusty CSM of the SN progenitor system that also gave rise
to the LBV-like outburst in 2004. We conclude that the flux difference was 
due to diffuse, irregular background emission. A similar excess was found
in the day~430 5.8~$\mu$m unsubtracted image. However, at 3.6~$\mu$m
the excess was about 50\%.  The fluxes obtained from the subtracted
images were adopted as the true SN fluxes.  At 4.5 and 8.0~$\mu$m the
true fluxes were estimated by scaling downward the values from the
unsubtracted images. The shift was 5\% for all except the second epoch
4.5~$\mu$m observation where we imposed a shift of 25\%, this being a
rough interpolation between the adjacent band shifts. The resulting
MIR fluxes are listed in Table~1.
\begin{table*}
\begin{center}
\caption{UKIRT WFCAM (days 86 - 231), UKIRT UFTI (day 279) and Gemini NIRI
(day 493) NIR magnitudes and Spitzer IRAC (days 228 and 430) MIR fluxes 
of SN 2006jc. The SN epochs (rounded to the nearest whole 
number) are relative to the estimated explosion date, JD = 2454000 (Pastorello et al. 
2008a). Uncertainties are shown in brackets. For completeness, we also tabulate 
the NIR photometry of Arkharov et al. (2006) used in our analysis. We note that more 
recently Di Carlo et al. (2008) have also reported updated photometry based on the dataset 
used by Arkharov et al.}
\begin{tabular}{lllllll}
\hline
Date (UT)     & JD-       &  Epoch &    J        &     H       &      K & Source \\ 
              & 2400000   &  (days)&             &             &        & \\ \hline
2006 Nov 15   & 54054.55  &  55    & 15.87       & 15.64       & 15.01       & Arkharov et al. \\
2006 Nov 16   & 54055.55  &  56    &  ---        & 15.53       &  ---        & Arkharov et al. \\ 
2006 Nov 24   & 54063.55  &  64    & 15.93       & 15.47       & 14.64       & Arkharov et al. \\
2006 Dec 03   & 54072.54  &  73    & 15.92       & 15.08       & 14.29       & Arkharov et al. \\
2006 Dec 06   & 54075.54  &  76    & 15.88       & 15.01       & 14.20       & Arkharov et al. \\
2006 Dec 16   & 54085.97  &  86    & 15.83(0.01) & 14.76(0.01) & 13.87(0.01) & This work \\
2006 Dec 23   & 54093.00  &  93    & 16.01(0.01) & 14.86(0.01) & 13.91(0.01) & This work \\
2006 Dec 30   & 54099.88  & 100    & 16.28(0.01) & 15.04(0.01) & 14.00(0.01) & This work \\
2007 Jan 13   & 54113.92  & 114    & 16.85(0.02) & 15.43(0.01) & 14.27(0.01) & This work \\
2007 Jan 20   & 54121.00  & 121    & 17.08(0.02) & 15.63(0.01) & 14.40(0.01) & This work \\
2007 Mar 16   & 54175.93  & 176    & 19.25(0.15) & 17.23(0.04) & 15.56(0.03) & This work \\
2007 Apr 26   & 54216.73  & 217    & -           & 18.17(0.08) & 16.39(0.03) & This work \\
2007 May 10   & 54230.74  & 231    & -           & 18.49(0.12) & 16.67(0.05) & This work \\
2007 Jun 27   & 54278.74  & 279    & -           & -           & 17.60(0.10) & This work \\ 
2008 Jan 27   & 54492.97  & 493    & -           & -           & 21.64(0.40)  & This work \\ \hline
              &           &        & 3.6$\mu$m    & 4.5$\mu$m    & 5.8$\mu$m    & 8.0$\mu$m \\ \hline
2007 May 7    & 54227.54  & 228    & 506$\pm$3 $\mu$Jy & 632$\pm$7 $\mu$Jy & 727$\pm$7 $\mu$Jy & 707$\pm$11 $\mu$Jy\\
2007 Nov 25   & 54429.67  & 430    & 49$\pm$2 $\mu$Jy & 92$\pm$10 $\mu$Jy & 199$\pm$6 $\mu$Jy & 286$\pm$10 $\mu$Jy\\\hline
\end{tabular}
\end{center}
\end{table*}

\section{Analysis}
\subsection{Evidence for dust from the IR spectral energy distribution}
To explore the evidence for dust we make use of our IR photometry (see
Table~1) and the optical photometry of Pastorello et al. (2007,
2008a).  We use also the $JHK$ measurements of Arkharov et al. (2006)
which cover epochs 55-76 days (Table~1), when the NIR light curves of
SN 2006jc were still rising (we note that more recently Di Carlo et
al., 2008 have also reported photometry based on the Arkharov et
al. dataset). To take an initially neutral standpoint on the
interpretation, we have compared blackbodies (see Fig. 3) with the
optical to NIR spectral energy distribution (SED) at each epoch
between 55 and 231 days, for which at least H and K-band data were
available.

\begin{figure}
\begin{minipage}{85mm}
\includegraphics[width=85mm, angle=0, clip]
{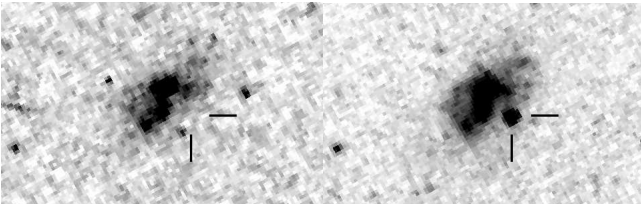}
\caption{The field of SN 2006jc at 5.8~$\mu$m. The FOV of the image is
107 $\times$ 77 arcsec and the SN location is marked with ticks.  The
images were obtained with IRAC on Spitzer on 2004 April 30 (left) and
2007 May 7 (right).  North is up and East to the left.}
\end{minipage}
\end{figure}

\begin{figure}
\begin{minipage}{85mm}
\includegraphics[width=85mm, angle=0, clip]
{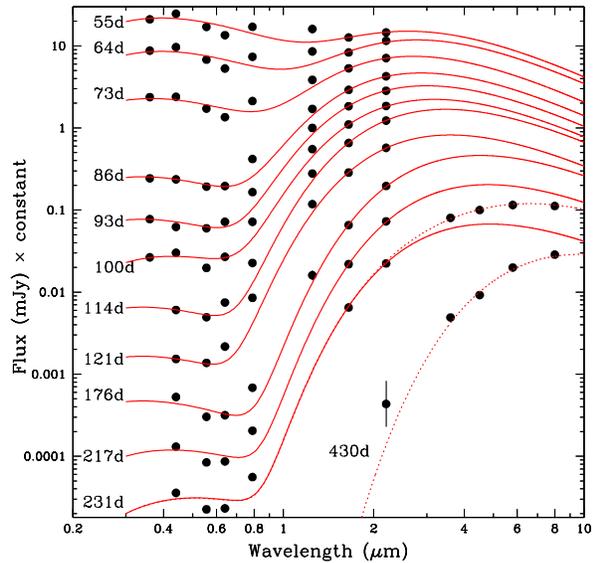}
\caption{Two component blackbodies compared with the optical - NIR
SEDs of SN 2006jc between 55 and 231 days (see Table~2). The
blackbodies have been reddened according to the Cardelli et al. (1989)
extinction law with $A_{\rm V} = 0.15$ (Pastorello et al. 2007).  The IR data
between 55 and 73 days are from Arkharov et al. (2006) and the rest of
the data are from this study. The optical data are from Pastorello et
al. (2007, 2008a) and have been interpolated to the epochs of the IR
observations.  The epochs are relative to the estimated explosion
epoch at JD = 2454000 (Pastorello et al. 2008a). The dotted lines show
that the MIR component at 231 days and the NIR and MIR components at
430 days can be reproduced by adding cool blackbodies (see Table~3).
The plots have been shifted vertically for clarity.}
\end{minipage}
\end{figure}
The optical photometric data were interpolated to the epochs of the
NIR observations.  The data at each epoch are compared with combined
hot and warm blackbodies. The parameter value evolution is presented
in Table 2 and Fig. 4. The optical part is represented by a hot
(10,000-15,000 K) blackbody, presumably due to the hot photosphere of
the SN, and the IR part by a warm (1050-1850~K) blackbody. The warm
blackbody is adjusted to optimise the match to the $HK$ points
only. This was done since contamination by line emission could be
relatively greater in the $J$ band. In practice, by
about 100~days the $J$ points were generally also well reproduced by
the $HK$-matched blackbodies.  The warm blackbody temperature stayed
around 1800 K for several weeks before declining to $\sim$1000~K by
day~231.  Its radius increased to about $0.8\times10^{16}$~cm by
day~176 and then declined.  This corresponds to a blackbody velocity
of $\sim$7000--8000 km/s between 55 and 140 days (see Fig. 4). The
luminosity of the warm component peaked around 90 days after which it
faded.  The contribution of the hot component was dominant at day 55
($>$90\% of the total luminosity) but by day $\sim$80, the warm
component luminosity exceeded that of the hot component and by day
217, the hot component was less than 1\% of the warm component.\\

A single blackbody was unable to reproduce both the NIR fluxes
at 231~days and the MIR fluxes at 228~days. It is unlikely that this
was due to the slightly different epochs. We therefore added a third
(cold) blackbody component to account for the MIR fluxes, and the
warm+cold blackbody combination is illustrated in Fig.~3 as a dotted
line.  A single cold blackbody reproduced the day~430 NIR/MIR SED
(Fig.~3) where the $K$-band point was obtained by interpolation
between the days 279 and 493 observations.  Owing to the uncertainty
in this procedure an error of $\pm0.7$ mags. was assigned to the
interpolated point. The parameter values for the days~228/31 and 430
warm and cold blackbodies are given in Table~3.  On day~228/31 the
luminosity of the warm component exceeded that of the cold by a factor
of $\sim$2. \\
\begin{table*}
\begin{center}
\caption{Parameter values of the hot and warm blackbodies matched to
the optical to the $H$ and $K$ fluxes of SN~2006jc}
\begin{tabular}{lllllllll}
\hline
Epoch      & radius& Temp.  & Lum. & radius & Temp.& Lum.           & Lum.             & $L_{hot}/L_{tot}$\\
(days)     & dust  & dust   & dust            & hot   & hot  & hot.           & tot.             & \\
  & ($10^{16}$cm)& (K)& ($10^{40}$erg/s) &($10^{14}$cm) & (K)  & ($10^{40}$erg/s)& ($10^{40}$erg/s)& (\%)\\     
\hline		   		     				   
55         & 0.34  &  1750  &   7.82& 2.18  &13000& 97.3&   105.1 & 92.6\\
64         & 0.42  &  1740  &  11.80& 1.82  &13000& 67.8&    79.6 & 85.2\\ 
73         & 0.47  &  1850  &  18.33& 1.26  &14000& 43.6&    61.9 & 70.4\\
76         & 0.51  &  1820  &  20.00& 1.18  &14000& 38.2&    58.2 & 65.6\\
86         & 0.63  &  1770  &  27.57& 0.59  &15000& 12.7&    40.3 & 31.5\\ 
93         & 0.67  &  1700  &  26.79& 0.39  &15000&  5.59&    32.4 & 17.3\\
100        & 0.72  &  1600  &  24.31& 0.39  &12000&  2.23&   26.5 &  8.4\\
114        & 0.77  &  1470  &  19.89& 0.15  &15000&  0.79&   20.7 &  3.8\\
121        & 0.78  &  1420  &  17.81& 0.10  &15000&  0.38&   18.2 &  2.1\\
176        & 0.83  &  1130  &   7.98& 0.055 &15000&  0.108&   8.09&  1.3\\
217        & 0.67  &  1065  &   4.18& 0.043 &12000&  0.027&   4.21&  0.64\\ 
231        & 0.63  &  1050  &   3.44& 0.046 &10000&  0.015&   3.44&  0.44\\
\hline
\end{tabular}
\end{center}
\end{table*}

\begin{figure*}
\begin{center}
\begin{minipage}{120mm}
\includegraphics[width=120mm, angle=0, clip]{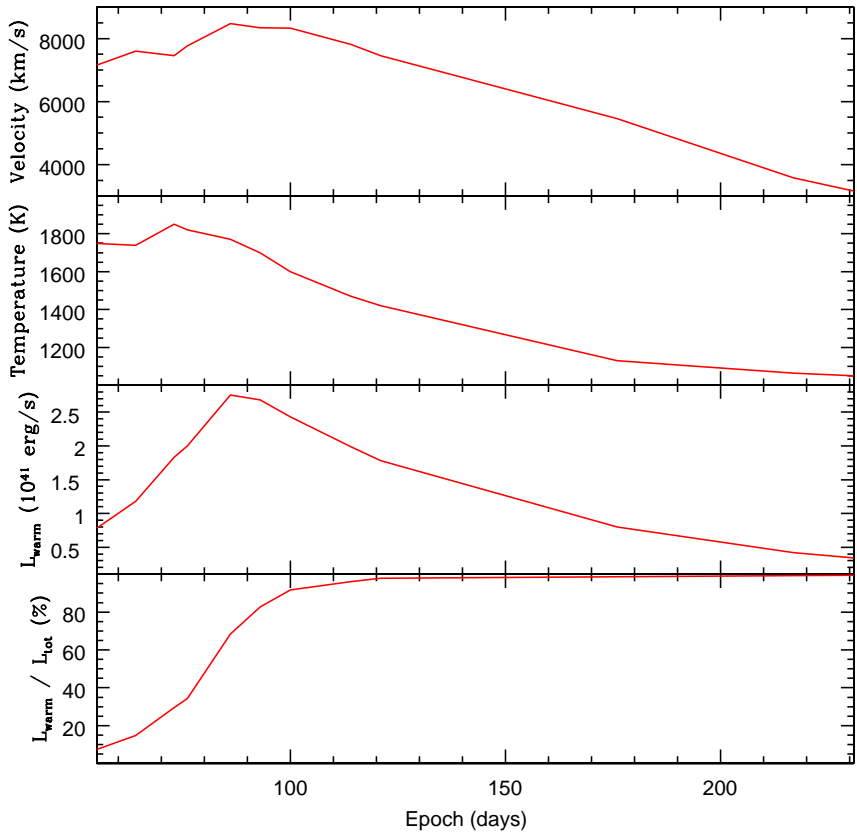}
\caption{Evolution of the parameter values of the warm
blackbodies matched to the $H$ and $K$ fluxes of
SN~2006jc (Table~2).  The blackbody velocities were obtained by
dividing the blackbody radii by the epochs of the observations.}
\end{minipage}
\end{center}
\end{figure*}

\begin{table*}
\begin{center}
\caption{Parameter values of the warm and cold blackbody matches to the days 228/31 
and 430 SEDs (see Fig. 3).}
\begin{tabular}{llll}
\hline
           & Temperature & Radius                & Luminosity   \\ \hline
Day 228/31 &&&\\
Warm blackbody  & 1050 K   & 0.63 $\times$ 10$^{16}$ cm & 3.44 $\times$ 10$^{40}$ erg/s \\
Cold blackbody  & 620 K    & 1.34 $\times$ 10$^{16}$ cm & 1.89 $\times$ 10$^{40}$ erg/s \\ \hline
Total lum.      &             &                         & 5.33 $\times$ 10$^{40}$ erg/s \\ \hline\hline
Day 430 &&& \\
Cold blackbody   & 520 K    & 1.5 $\times$ 10$^{16}$ cm & 1.17 $\times$ 10$^{40}$ erg/s \\ \hline
\end{tabular}
\end{center}
\end{table*}

Given the temperatures, sizes and luminosities of the warm and cold
blackbodies plus the evolution of the warm component, the most
plausible explanation for these components of the SED is thermal
emission from dust in the SN ejecta and/or in the surrounding medium.
A similar conclusion was reached by Smith et al. (2008), as well
as by Sakon et al (2008), Tominaga et al. (2008), Di Carlo et al
(2008) and Nozawa et al. (2008).  Further interpretation requires us
to address the location and energy source of the radiating dust. We
shall consider IR emission from newly-formed dust in the ejecta and/or
in a shell formed by the interaction of the SN ejecta with
circumstellar material. We shall also consider emission from
pre-existing dust in the progenitor wind.

\subsection{Evidence for new dust from line profiles and light curves}
We sought evidence of newly-formed dust via the optical line profiles.
Figure~5 illustrates the evolution of the profiles of the narrow He I
and H$\alpha$ lines using data from Pastorello et al. (2007, 2008a).
The He I lines show a clear blueshift which increased to $\sim$600
km/s between $\sim$60 and $\sim$100 days. Over the same period,
Smith et al. (2008) also found the He I lines becoming progressively
more asymmetric and blueshifted. Concurrently, the width of
the line profile decreased.  The FWHM of the He I $\lambda$7065 line
decreased from $\sim$2400 km/s to $\sim$1800 km/s between 54 and 96
days.  There is less-convincing evidence for a blueshift in the narrow
H$\alpha$ line. There may be a modest shift between 84 and 140~days
but the signal-to-noise is low. Over the same period, Smith et
al. (2008) find that the H$\alpha$ profile is not systematically
shifted to the blue.  Blueshifts observed in broad SN lines e.g. in
the case of SN 1987A, have been attributed to dust condensation in the
SN ejecta (e.g., Danziger et al. 1989; Lucy et al. 1989).  Similar
evidence for dust has been reported for SNe~1999em and 1990I (Elmhamdi
et al. 2003, 2004), SN~1998S (Pozzo et al. 2004) and SN 2003gd
(Sugerman et al. 2006).  However, such
blueshifts were not observed in the broad lines of SN~2006jc. Instead,
the broad lines simply disappeared by 140 days (see Fig. 5).  In
contrast, the narrow He I lines persisted until at least 182 days.  A
detection of He~II~$\lambda$4686 emission was reported by
Smith et al. (2008), appearing some time between days 71 and 95 and
disappearing between 122 and 148 days.  Our closest observation epochs
to these were at 88 and 132~days but no sign He~II~$\lambda$4686
emission was found. It is conceivable that our observations did not
cover the period when the He~II emission was strong.  In Figure~6 we
compare the evolution of the He~I~$\lambda$7065 line center in
velocity space (top panel) with the optical and NIR light curves of
the SN (middle panel).
\begin{figure}
\begin{minipage}{85mm}
\includegraphics[width=42mm, angle=0, clip]{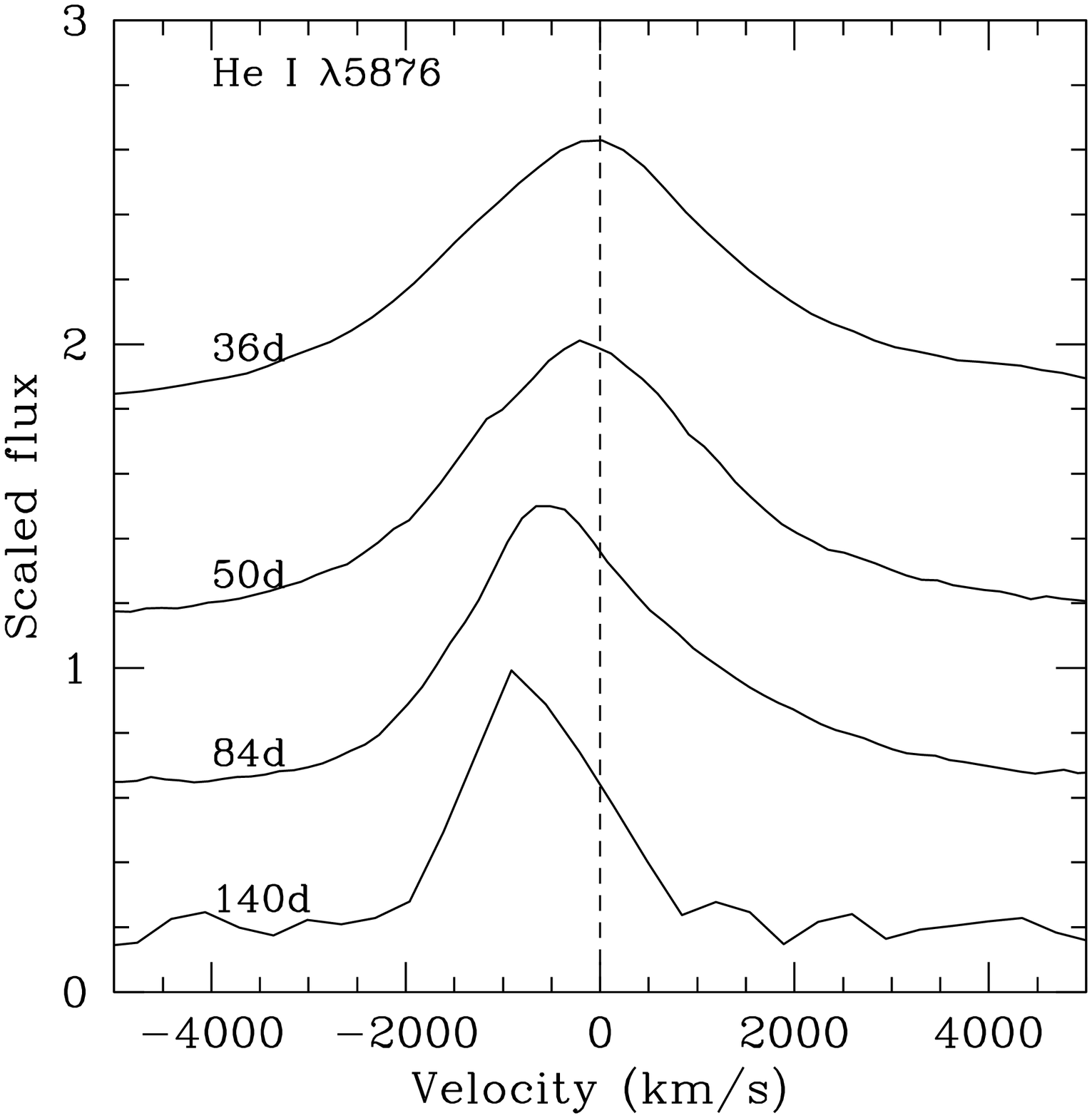}
\includegraphics[width=42mm, angle=0, clip]{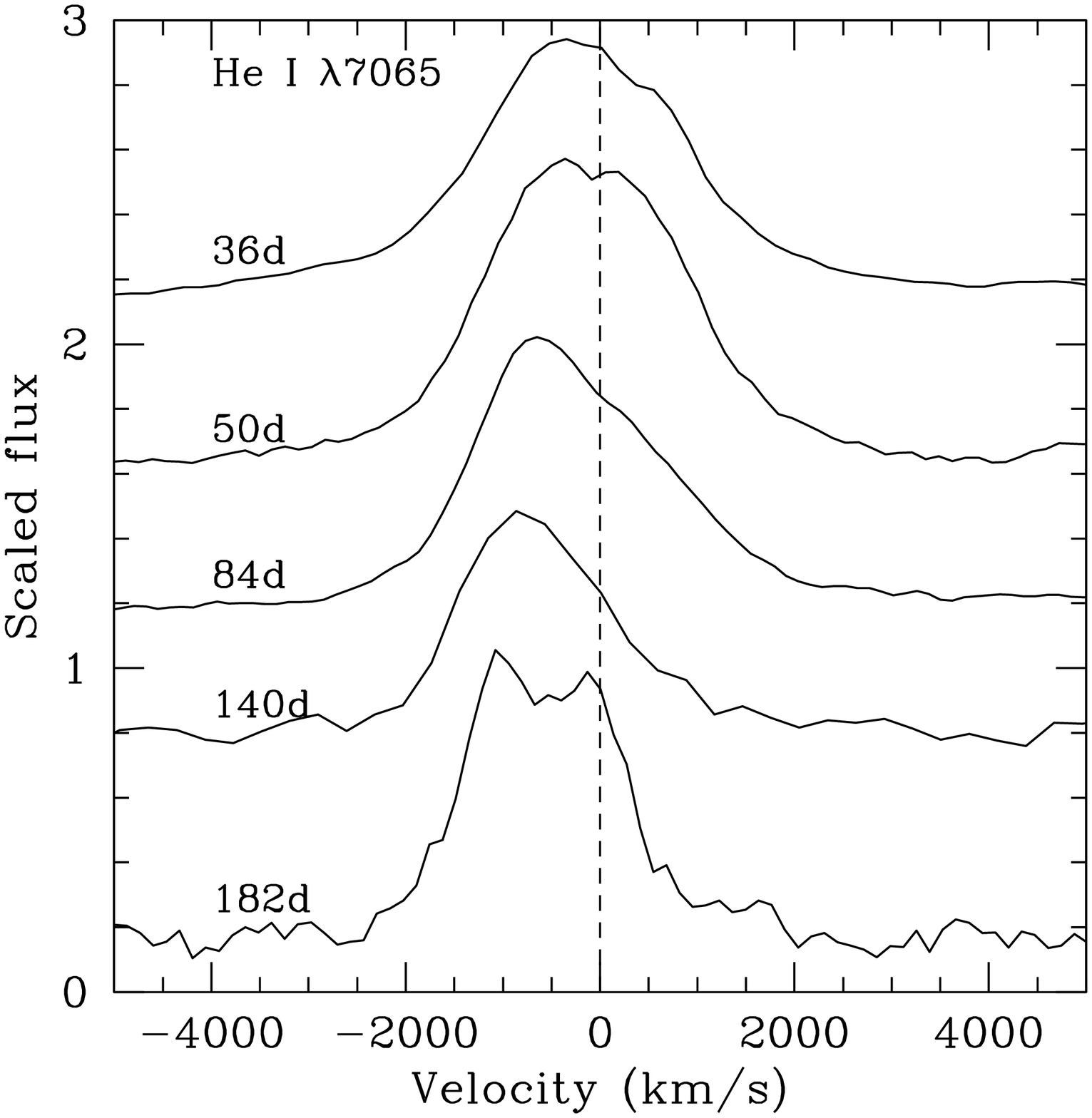}
\includegraphics[width=42mm, angle=0, clip]{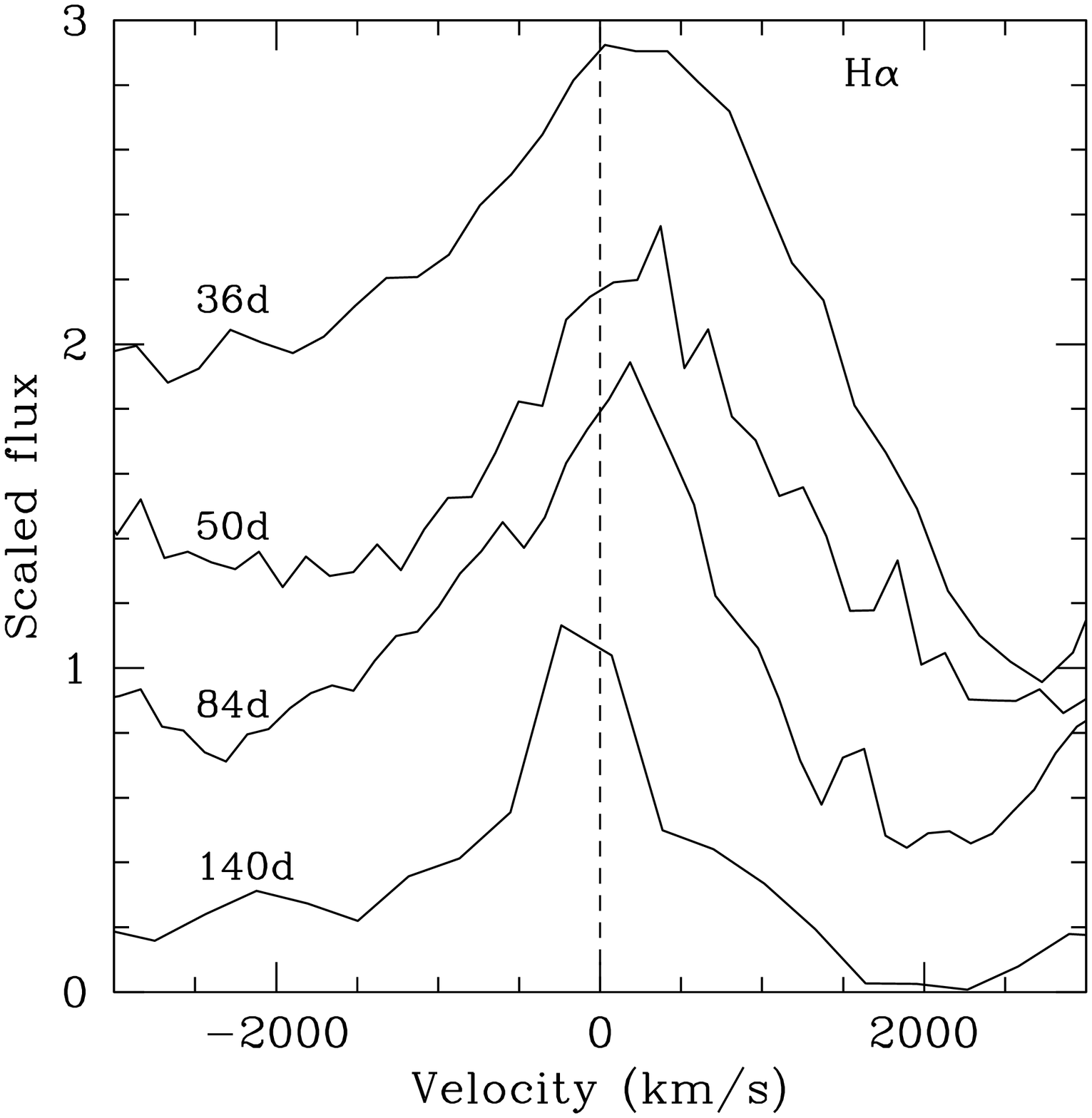}
\includegraphics[width=42mm, angle=0, clip]{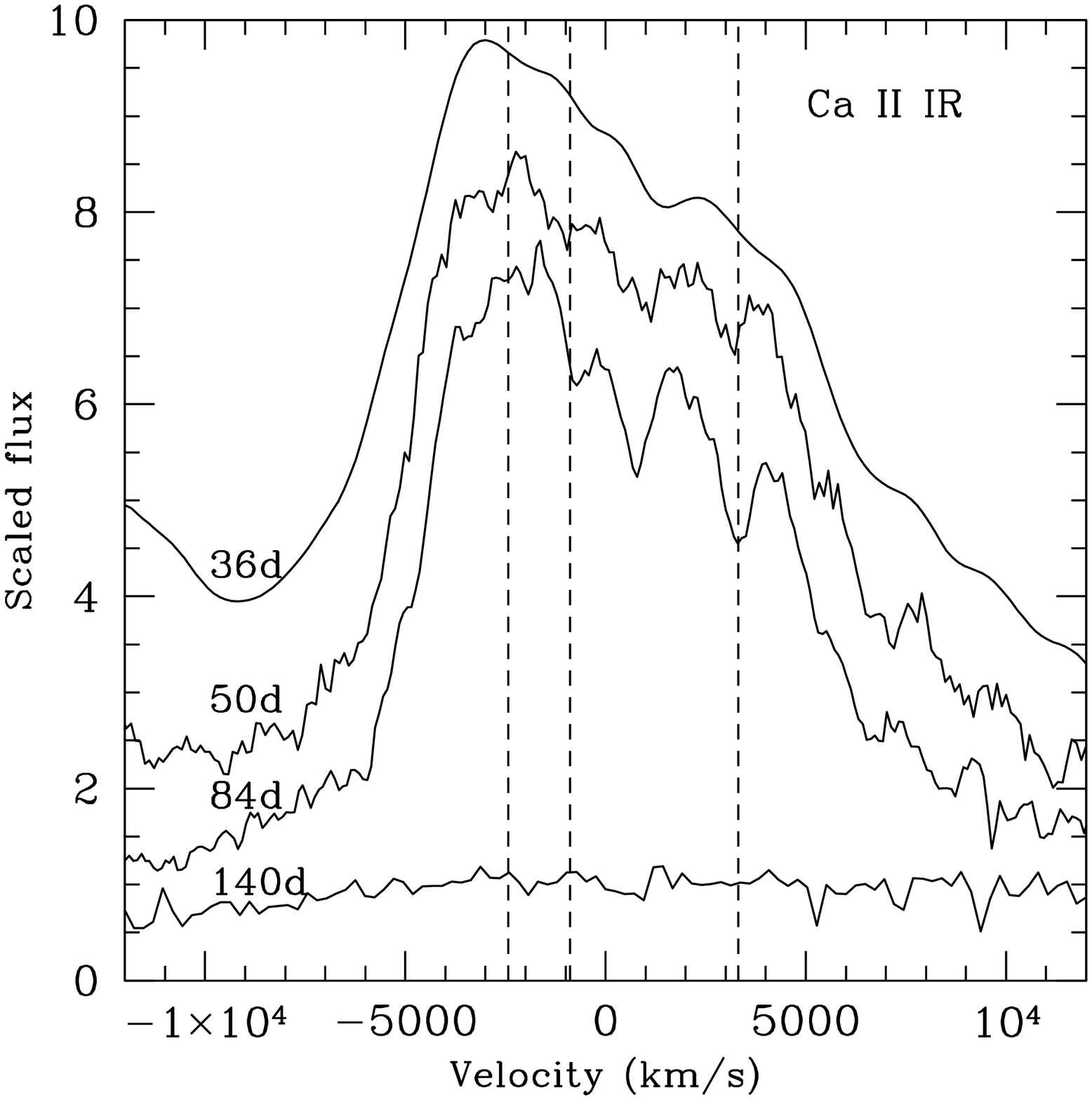}
\caption{Evolution of the spectral line profiles of He I $\lambda$5876
and $\lambda$7065, and H$\alpha$ in velocity space. The evolution of
the Ca~II IR triplet profile is also illustrated. The velocities are
w.r.t. the average wavelength of the three lines. The zero velocities
(w.r.t.  the host galaxy recession velocity) are marked
by vertical dashed lines.}
\end{minipage}
\end{figure}
It can be seen that the He~I line blueshift started to develop at about the 
same time as the NIR excess appeared.  This supports the
proposal of Smith et al (2008) (see also Sakon et al. (2008), Tominaga et al. (2008),
Di Carlo et al. (2008) and Nozawa et al. (2008)) that the NIR excess 
is indeed due to local dust condensation. We also note that, as Smith et al. 
point out, the Ca II triplet line (Fig.~5) simply fades, and does not show the
characteristic blueshift of He I lines.\\

We also examined the individual light curves for indications of dust
formation.  At about 65 days (10 days after the first evidence of an
IR excess) an abrupt steepening of the $UVBRI$ light curves can be
seen. The steeper slopes persist until about 120 days. The steepening
becomes more pronounced as we move to shorter wavelengths, although
this reddening is not a strong effect. Relative to the slopes seen in
the period 50 to 65 days, the additional attenuation by day 120 is
quite substantial viz. roughly $A_B=3.0$, $A_V=2.8$, $A_R=2.8$,
$A_I=1.8$.  We propose that the slope steepening is due to attenuation
by newly-formed dust.  The lack of strong wavelength dependence may be
due to the dust forming in clumps rather than in a uniform
distribution.\\

We have also created optical and optical+NIR 'quasi-bolometric' light
curves (quasi-BLCs) by integrating over the SN SED between $U$ and $I$
band and $U$ and $K$ band, respectively, at each epoch (see Pastorello
et al. 2008a for details).  Zero flux was assumed at the blue edge of
the $U$ band and at the red edge of the $I$ or $K$ band.  It is
important to note that this procedure implicitly assumes that the
optical and NIR fluxes have contemporary energy sources.  If a
significant fraction of the NIR arose from an IR~echo then, owing to
light travel time effects, simple addition of the optical and IR
fluxes might not give a meaningful quasi-BLC unless the distance of
the re-radiating dust from the ejecta was small enough (see
Sect. 4.4 and 4.5).  The quasi-BLCs are shown in the bottom panel
of Figure~6 as open and filled stars, respectively.  The optical
quasi-BLC exhibits a steepening relative to the optical+NIR BLC, thus
supporting the dust condensation hypothesis.  It might be argued that,
since the $U$ to $K$ BLC does not include photospheric emission
shortward of the $U$ band nor dust emission longward of the $K$ band,
it does not give a true picture of the supernova's bolometric
evolution (see also Pastorello et al. 2008a).  We therefore created a
more realistic BLC by summing the luminosities of the hot and warm
blackbody components at each epoch (see Table 2). The resulting BLC is
shown in Fig.~6, bottom panel. At 55~days, the blackbody-based BLC
luminosity exceeds that of the $U$ to $K$ BLC by about $\times$2,
rising to $\times$6 by 200~days. At early times this is due to the
contribution of the unobserved UV emission shortward of the $U$-band
and at later times the growing IR contribution beyond the $K$ band
(see Fig.~3). We note that the optical quasi-BLC also shows steepening
relative to this blackbody-based BLC. We conclude that the light
curves provide evidence of dust formation in the SN~2006jc vicinity.
For comparison the $U$ to $M$ band BLC of SN~1987A (Suntzeff \&
Bouchet 1990) is also shown in Fig.~6.  We note that the post-130 day
BLC of SN 2006jc has roughly a third of the luminosity of SN 1987A's
BLC. This is in spite of the estimated $^{56}$Ni mass (Tominaga et al. 2008;
Pastorello et al. 2008a) being $\times$3-6 greater. The explanation for this 
is the (possibly) much lower ejecta mass, together with the higher
($\sim\times5$) ejecta velocity of SN~2006jc. Consequently the
gamma-ray transparency of the SN~2006jc ejecta increases much more
rapidly.  The BLCs of SN 2006jc will be used in Sect. 4.4-4.7 as input
luminosities for IR~echo models.  The day~50 epoch is indicated in
each panel as a vertical line. This is the approximate date at which
we deduce that dust condensation commenced (see below).

\begin{figure*}
\begin{minipage}{160mm}
\center{
\includegraphics[width=120mm, angle=0, clip]
{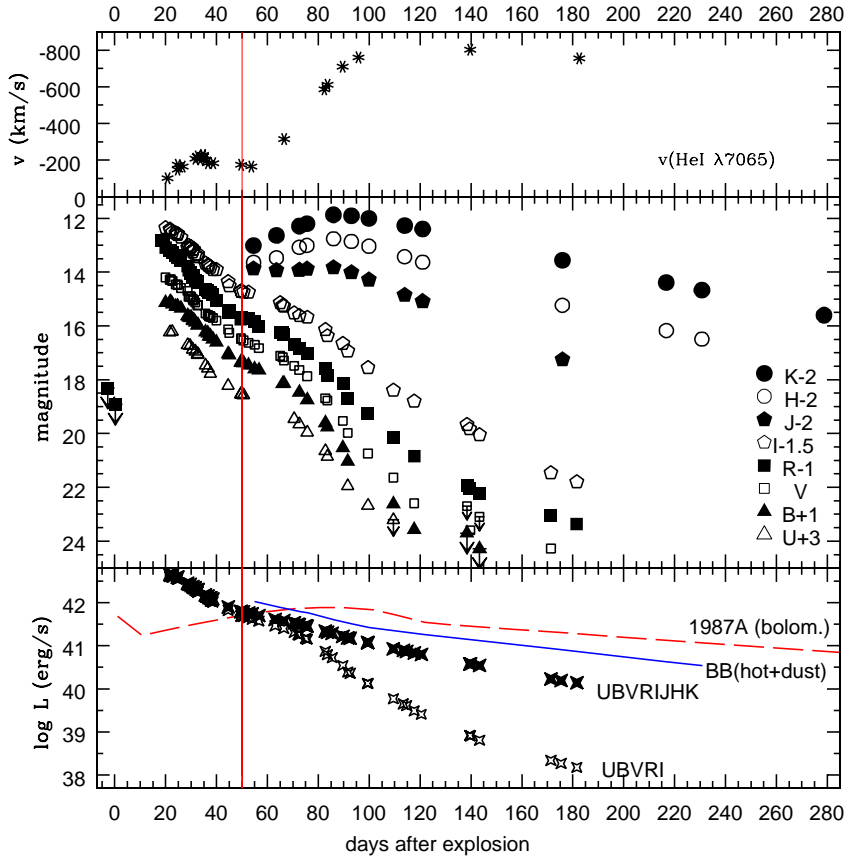}}
\caption{Top: The velocity of the line center of He I $\lambda$7065
with respect to the rest wavelength as a function of the SN epoch
(data are from Pastorello et al. 2007, 2008b).  Middle: Optical
(Pastorello et al. 2007, 2008a) and NIR light curves of SN 2006jc.
Bottom: Optical (U to I) and optical+NIR (U to K) quasi-bolometric
light curves of SN 2006jc (Pastorello et al. 2008a). Also shown is the
bolometric light curve of SN~2006jc obtained by summing the hot and
warm component blackbody luminosities at each epoch (Table 2). In
addition the bolometric (U to M band) light curve of SN 1987A
(Suntzeff \& Bouchet 1990) is plotted for comparison.  The day~50
epoch is indicated in each panel as a vertical line. This is the
approximate date at which we deduce that dust condensation commenced
(see text).  }
\end{minipage}
\end{figure*}

\subsection{Newly-formed dust in the ejecta?}
For newly-formed ejecta dust to attain the warm blackbody radii given
in Table~2, the blackbody radius must have expanded at about 8000 km/s
up to $\sim$120 days (see also Fig.~4).  A similarly high blackbody velocity 
($\sim$7000 km/s) is required for the cold component in the 3-blackbody 
match at 228/31 days (see Table 3).  In more typical CCSNe, such velocities
would imply that much of the IR emission could not have been due to
newly-formed dust in the SN ejecta as there are not enough suitable
refractory elements available for dust condensation at such high
velocities.  For example, the nebular spectrum of the Type~IIP
SN~2003gd indicated that the bulk of the metals lay at velocities
below $\sim$2000~km~s$^{-1}$ (Hendry et al. 2005).  However, higher
metal velocities are found in other SN types.  In particular, Type~Ic
supernovae exhibit nebular metal velocities of 5000-8000~km/s
(Filippenko et al.  1995; Taubenberger et al. 2006).  Despite the fact
that the latest spectrum available for SN~2006jc (183~days)
(Pastorello et al. 2008a) shows no sign of nebular metal lines,
the probable origin of the SN in a star lying on the Ib/Ic progenitor
boundary suggests that such high velocities might also be present
here.\\

To explore further the possibility that new ejecta dust was the source
of the SN~2006jc NIR luminosity, we matched a simple dust IR-emission
model to the observed fluxes. The model is based on the escape
probability formalism (Osterbrock 1989; Lucy et al. 1989; Meikle et
al.  2007), in a spherical configuration.  (The MIR component will be
considered later.) An additional component was added to represent
continuum emission from hot, optically thick gas in the
ejecta. Details of this model are given in Meikle et al. (2007).  We
adopted a uniform dust density and temperature and investigated dust
comprising either (a) pure amorphous carbon, or (b) pure silicates.
The mass absorption functions were taken from the literature (Rouleau
\& Martin 1991; Laor \& Draine 1993).  The grain size distribution law
was set at $m=3.5$ (Mathis et al. 1977).  The distribution limits were
set at $a_{min}=0.005~\mu$m and $a_{max}=0.05~\mu$m, these being based
on the typical grain size ranges calculated by Todini \& Ferrara
(2001) and Nozawa et al. (2003). The free parameters were the grain
temperature, radius of the sphere, and grain number density scaling
factor.  The model results are listed in Table~4.  Between 55 and
121~days the sphere remained optically thick and expanded at a
constant velocity of about $8500\pm500$~km/s. After this time, the
observed fluxes could be reproduced by either reducing the expansion
velocity or by allowing the dust to become optically thin. We suggest
that it is more likely that newly-formed ejecta dust would expand
steadily and so we fixed the expansion velocity at 8500~km/s for the
latest three epochs.  \\

\begin{table}
\caption{Dust sphere model results for newly-formed dust in the ejecta.}
\begin{tabular}{rlllrl}
\hline
Epoch   & M$_{dust}$ &  $\tau_{2.2mic.}$ & Radius & Temp.\\
(days)  & ($10^{-4}$~M$_{\odot}$)&        & ($10^{16}$cm)  &  (K) \\
\hline
Am. Carbon &&&&& \\
\hline
55      &   0.10  &    1.82 & 0.36 &  1800    \\
64      &   0.23  &    2.40 & 0.48 &  1700    \\
73      &   0.29  &    2.59 & 0.52 &  1800    \\
76      &   0.30  &    2.44 & 0.55 &  1800    \\
86      &   0.47  &    2.62 & 0.66 &  1760    \\
93      &   0.60  &    2.85 & 0.72 &  1670    \\
100     &   0.68  &    2.71 & 0.78 &  1570    \\
114     &   0.67  &    2.44 & 0.82 &  1450    \\
121     &   0.88  &    2.67 & 0.89 &  1370    \\
176     &   0.62  &    0.90 & 1.29 &  1050    \\
217     &   0.42  &    0.39 & 1.59 &   950    \\
231     &   0.28  &    0.24 & 1.70 &   950    \\
\hline \hline
Silicates &&&& \\
\hline
55      &   2.0  &    2.6  & 0.33 & 1850    \\
64      &   2.5  &    2.6  & 0.51 & 1650    \\
73      &   2.4  &    2.4  & 0.52 & 1800    \\
76      &   3.0  &    2.4  & 0.55 & 1800    \\
86      &   3.9  &    2.4  & 0.66 & 1760    \\
93      &   4.6  &    2.4  & 0.73 & 1670    \\
100     &   5.4  &    2.4  & 0.79 & 1570    \\
114     &   6.0  &    2.4  & 0.84 & 1450    \\
121     &   7.3  &    2.5  & 0.89 & 1370    \\
176     &   5.3  &    0.88 & 1.29 & 1050    \\
217     &   2.7  &    0.29 & 1.59 &  980    \\
231     &   2.4  &    0.23 & 1.70 &  950    \\
\hline
\end{tabular}
\end{table}

For both grain materials, to match the $HK$ fluxes up to about
100~days required a dust sphere expansion velocity of 7000-9000~km/s
and also that the dust was optically thick at wavelengths up to at
least 2.2~$\mu$m (Table~4). In practice we set the optical depth in
the $K$ band at about 2.5. For a Cardelli et al. (1989)
extinction law with $R_V=3.1$ this corresponds to $A_V=22$.  However,
to achieve a match in the 55-100~day period required a temperature of
1600~K--1850~K for both amorphous carbon and silicate dust.  Such high
temperatures immediately ruled out silicates as the dust material
(Smith et al. (2008) reached the same conclusion).  For example,
Todini \& Ferrara (2001) find that while amorphous carbon grains form
in the temperature range 1650--1900~K, for silicate grains the
temperature must fall to 1100--1300~K before condensation occurs.
This is consistent with the absence of the silicate feature in the
vicinity of 8~$\mu$m at 228 and 430~days (see Fig~3). Henceforth, we
therefore focus our attention on amorphous carbon grains.  Assuming
that the expansion velocity remained at $\sim$8500~km/s, we found that
the dust optical depth declined significantly by 176~days becoming
optically thin (at 2.2$\mu$m) by day~217.  After 100~days, the
temperature falls from $\sim$1500~K to about 950~K by 231~days (see
Table 4).  The dust mass grew to $\sim10^{-4}$~M$_{\odot}$ by day~121
and then declined to a third of this value by 231~days. The apparent
decline in mass may be due to some dust cooling to below detectability
in the $HK$ bands.  We conclude that a uniform amorphous carbon dust
sphere can plausibly reproduce the NIR fluxes. As already indicated,
the maximum expansion velocity is in line with the velocities seen in
Type~Ic nebular metal spectra. The maximum temperature is reasonable
for amorphous carbon grain precipitation and the dust mass is
modest.\\

The high optical depth might be seen as a problem as it would totally
block out the optical emission from the ejecta. This seems to conflict
with the much smaller estimated extinction (see above), but clumping
of the ejecta could allow a sufficient fraction of the flux to escape
to yield consistency with the extinction.  One possible difficulty is
the extraordinarily early appearance of the dust viz. $t\sim50$~days.
This contrasts with the well-studied SN~1987A where the earliest
evidence of dust formation was at $\sim$350~days post-explosion
(Meikle et al. 1993).  This is consistent with the dust condensation
calculations of Todini \& Ferrara (2001) who found that the earliest
dust (amorphous carbon) would appear at about 1 year. However, recent
calculations by Nozawa et al. (2008) suggest that such early dust
condensation is possible in SN~2006jc (but see discussion in
Sect.  5).  A more serious difficulty is how to account for the
attenuation of the narrow He~I line red wings.  If we accept that
these lines are indeed due to a shell of material ejected at
$\sim$2400~km/s in the LBV-like outburst of October 2004 then by epoch
121~days, when the maximum line shift was attained (see Fig.~6), the
projected area of the putative ejecta dust sphere would be only
$\sim$30\% of the projected shell and so would attenuate only the
reddest 15\% of the red wing. In fact, almost the entire red wing had
vanished by 140~days (see Fig.~5). Furthermore, there was no evidence
for attenuation of the red wings of the broad SN lines that would have
been expected if dust had formed within the ejecta. A similar argument
has been made by Smith et al. (2008).  We conclude that, in spite of
the success of the ejecta dust sphere model in accounting for the
observed IR emission, the spectral evidence argues against significant
dust condensation in the ejecta. \\

\subsection{An IR~echo from pre-existing CSM dust?}
We have argued that the IR behaviour of SN~2006jc strongly indicates
the presence of dust in the supernova vicinity. However, the strong
blueshifts in the He~I lines as well as, perhaps, the large radii of
the blackbodies and the early appearance of the IR excess emission
argues against newly-formed dust in the ejecta.  Nevertheless, the
evolution of the He~I spectral profiles and the behaviour of the
individual light curves and quasi-BLCs (see Fig. 6) point to dust
condensation taking place during an approximate 2--4 month period
after the explosion. However, such dust may or may not also be
responsible for the IR luminosity. We therefore first examine the
latter possibility.  Given the evidence that the progenitor of
SN~2006jc was a massive, highly-evolved star, we are prompted to
explore the possibility that the bulk of the IR emission arose from
pre-existing dust in the progenitor wind heated primarily by the
early-time UV/optical emission from the SN i.e. an IR~echo. \\

The model used to test the IR~echo hypothesis follows those of Bode \&
Evans (1980), Dwek (1983), Graham \& Meikle (1986) and Meikle et
al. (2006).  A spherically-symmetric dust cloud having a single grain
size is adopted, with the actual value of the grain radius as a free
parameter.  UV/optical absorption and IR emission for the grains is
calculated realistically using the emissivity function for amorphous
carbon. Silicates are dismissed for the reasons given above (also see
Dwek 1985).  The input luminosity was a parameterised description of
the $UBVRI$ bolometric light curve of Pastorello et al. (2008a).  To
allow for the unobserved flux shortward of the $U$-band, the
luminosity was scaled up by a factor of about $\times1.9$.  The
unobserved 0--22~day part was represented using a plausible
extrapolation of the $UBVRI$ BLC viz.
$L_{Bol}=1.2\times10^{43}e^{-t/84.4(d)}$~erg/s (including the
$\times1.9$ scaling).  The later-time IR excess light curves were not
included in the input BLC for the following reason.  For an IR~echo
from the pre-existing dust in the progenitor wind, owing to
light-travel time, the resulting later-time IR excess light curves
would be dominated by the SN UV/optical luminosity around the time of
peak emission. Valid addition of the IR light curves would therefore
require them to be de-convolved from the light-travel time delays
introduced by the echo process and the size of the dust-free cavity.
We set the dust-free cavity radius to be $8\times10^{16}$~cm (30~light
days), this being the distance from the SN at which the dust only just
reaches the approximate evaporation temperature for amorphous carbon
grains of 1800~K. For this size of cavity, the contribution of the
appropriately de-convolved IR light curve to the UV/optical light
curve would be small and so no attempt was made to include this small
contribution to the BLC.  The outer radius of the CSM was set at
$10^{18}$~cm, although the actual value is not critical.  
We varied the grain radius, grain number density, CSM radial
density law and the adopted characteristic wavelength of the input
UV/optical radiation until the observed NIR SED and its evolution were
reasonably reproduced. A grain radius of $a=0.05\mu$m, a $r^{-2.25}$
density law, and a characteristic input wavelength of 0.3~$\mu$m were
settled upon.  The dust mass is a modest
0.27$\times$10$^{-3}$~M$_{\odot}$, corresponding to a plausible CSM
mass of $0.027 \times (0.01/r_{dg})$~M$_{\odot}$ where $r_{dg}$ is the
dust-to-gas mass ratio.  The UV/optical optical depth is low ($\tau$
$\sim$ 0.056) and so would not significantly attenuate the SN flux, in
agreement with the low early-time extinction towards SN~2006jc of
$A_{\rm V}\leq$0.15 (Pastorello et al. 2007).  However, in order to match
the observed NIR SED, it was also necessary to increase the input BLC
luminosity by a further factor of $\times3$. Even taking into account
the uncertainties in the SN distance and in the characteristic photon
wavelength of the early-time BLC, plus the possibility of a
contribution to the input luminosity by the IR component of the BLC at
later times, such a factor does seem rather large.  The need for the
introduction of this factor suggests that a simple progenitor wind
echo cannot account for the IR behaviour of SN~2006jc. Further
evidence against this scenario is given below.  In Figure~7 we compare
the echo model matches with the $H$ and $K$-band light curves (LH
panel) and the 100~day and 228/31~day IR SEDs (RH panel).
\begin{figure}
\begin{minipage}{85mm}
\includegraphics[width=85mm, angle=0, clip]
{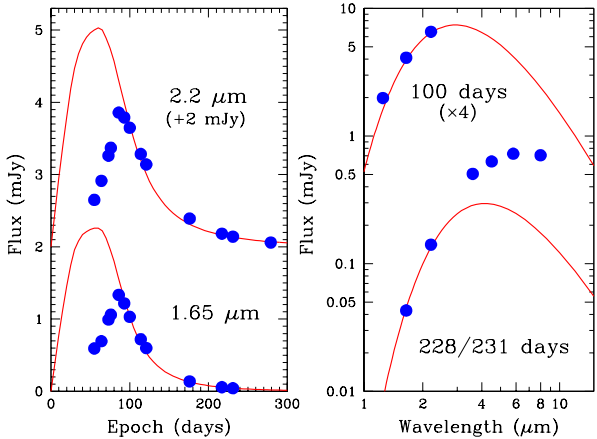}
\caption{CSM wind IR~echo model compared with the $H$ and $K$-band
light curves of SN 2006jc (LH panel), and with the 100~day NIR SED and
the 228/31~day MIR SED (RH panel). The $K$-band light curve and the
100~day SED have been shifted vertically for clarity.  The model
parameters and their values are as follows. Grain type: amorphous
carbon, grain material density: 1.85~g~cm$^{-3}$, grain radius: 0.05
$\mu$m, emissivity law: $\lambda^{-1.15}$, CSM density law:
$r^{-2.25}$, $r_{in}=8\times10^{16}$~cm, $r_{out}=10^{18}$~cm,
$\tau_{UV/optical}$=0.056, dust mass: $0.27\times10^{-3}$~M$_{\odot}$,
CSM mass: $0.027 \times (0.01/r_{dg})$~M$_{\odot}$, distance:
25.8~Mpc. To achieve the matches, the input SN BLC luminosity has been
multiplied by $\times$3.  Also, note the failure of the model to
reproduce (a) the delayed rise in the light curves, and (b) the strong
late-time MIR SED.}
\end{minipage}
\end{figure}
The NIR SEDs and the downward parts of the NIR light curves are well
reproduced by the echo model. However, in addition to the rather
implausible upward scaling of the input BLC, two other difficulties
are apparent.  One is that the model severely underproduces the MIR
SED on day~228.  The other problem is that the delayed rise in the IR
light curve is not reproduced. While such a delay might be generated
by placing the bulk of the dust on the far side of the SN, we
regard this as an unattractive {\it ad hoc} solution.\\

We conclude that the pre-existing dust IR~echo hypothesis is a rather
implausible means of accounting for the overall IR behaviour of
SN~2006jc.  Moreover, such an explanation requires that the optical
attenuation effects in the spectra and UBVRI light curves are due to a
separate, newly-formed, dust population which condensed early and
quickly.  However, as shown in Section 4.7, such an IR~echo provides a
good explanation for a significant and increasing proportion of the IR
flux between 228 and 493~days.\\

\subsection{Newly-formed dust in the shocked CSM}
We now consider the possibility that the bulk of the NIR emission was
due to emission from dust lying much closer to the supernova than the
closest pre-existing dust that could have survived.  To escape
evaporation by the early-time SN luminosity, such dust would have to
form {\it after} the peak luminosity had passed. The dust would then
be able to condense within the dust-free zone surrounding the
supernova. Such a cavity could have been created by the SN peak
luminosity or by a low mass-loss rate period during episodic
progenitor mass loss.  The dust might be formed during the interaction
of the ejecta with either a steady wind from the SN progenitor star,
or with a dense shell of material ejected in a discrete event in the
progenitor's past.  It has been recognised for many years
(e.g. Chevalier 1982) that the interaction of the supernova with a
dense CSM produces forward and reverse shocks. When radiative cooling
is important at either shock front, the gas can undergo a thermal
instability forming a dense, relatively cool zone i.e. a cool dense
shell (CDS).  Pozzo et al. (2004) invoked the formation of a CDS
formed behind the reverse shock to explain the post-300~day IR excess
observed in the Type IIn SN~1998S.  In this case the CDS was composed
mostly of ejecta material. In contrast, in the forward shock case the
CDS forms primarily out of CSM material.  If we assume a steady
progenitor wind, the mass loss rate from the progenitor star would
need to be very high to produce a significant amount of dust.
However, if the outward shock encounters a pre-existing dense
circumstellar shell then, as pointed out also by Smith et al (2008),
substantial dust formation can take place with a much reduced net mass
loss from the progenitor.  The latter case is pertinent to our
SN~2006jc study since, as described above, a shell of circumstellar
material was probably ejected during the LBV-like outburst seen 2
years before the SN explosion.  We therefore adopt the second scenario
in the analysis that follows.  This is illustrated in Figure~8. As
mentioned above, a dense shell behind the outward shock has already
been invoked by Smith et al. (2008) and Di Carlo et al. (2008) as the
main location for dust condensation in SN~2006jc (see Sect. 5 for
further discussion).\\

\begin{figure}
\begin{minipage}{85mm}
\includegraphics[width=85mm, angle=0, clip]
{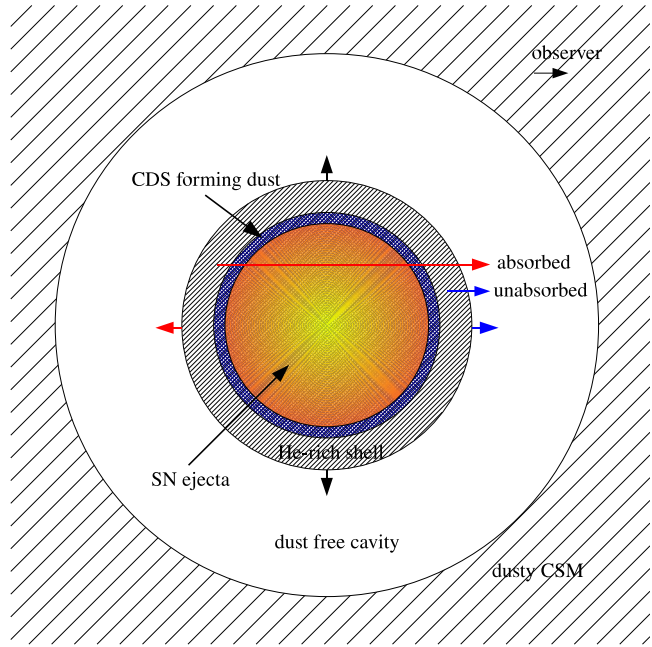}
\caption{ Schematic illustrating the geometry of the newly formed 
and pre-existing dust around SN 2006jc.  }
\end{minipage}
\end{figure}

To provide guidance for our interpretation of the IR luminosity within
the shell context, we modelled the shock evolution.  We propose that
the ejecta impacted on the He-rich shell ejected during the LBV-like
outburst in 2004 October and that a shock ran through the
shell. During a significant fraction of the interaction time, the
shock would have been radiative and dust could have condensed behind
the shock.  The LBV-like outburst occurred at about --730~days.  Let us
assume that the ejected shell expanded with a constant velocity of
$\sim$2400~km/s, as suggested by the He~I linewidths, up to the epoch
of earliest dust detection at 55~days.  We know that the duration of
the LBV-like outburst was no more than one month (see Supplementary
Table 2 in Pastorello et al. 2007), implying a shell thickness of
$\leq$4\% of the shell radius at the beginning of our observations.
The interaction was therefore modelled assuming a thin shell
approximation. While this is valid for the radiative forward shock,
the reverse shock propagating back into the SN ejecta will not be
radiative and so the formation of a significant CDS behind it is
unlikely.  Neither do we consider relativistic effects which could be
important for the very early interaction with ejecta moving at
velocities above $\sim$50,000~km/s.  The structure of the shocked CSM
shell was calculated using the same method as in Lundqvist \& Fransson
(1988), i.e. when the cooling timescales are short compared with the
hydrodynamical timescales, the steady state solution of the
hydrodynamical equations can be solved using standard numerical
techniques.  These equations take into account spherical
geometry. Ions and electrons are treated separately, with the energy
exchange between these particles specified as in Spitzer (1962).
Accurate radiative cooling is also calculated.  This is carried out
using a plasma code which calculates the ionization and emissivity as
a function of electron temperature. All ionization stages of the
elements were included, as well as all important types of emission
(i.e., free-free emission, recombination emission, two-photon emission
and line emission). The plasma code is described in some detail in
Sorokina et al. (2004).\\

Following Model~C of Pastorello et al. (2008a) and the models of
Tominaga et al. (2008), we assumed a SN explosion energy of
$\sim$$10^{52}$~ergs.  The mass of the ejecta and the power law index,
$n$, of the presumed $r^{-n}$ ejecta density profile of the outermost
ejecta, were varied between 4 and 10~M$_{\odot}$, and between 8 and
12, respectively. The fastest ejecta expand at velocities in excess of
$\sim$30,000~km/s and so would have reached the shell by 55~days when
the IR excess was first observed. The ejecta were collided with a
shell of CSM material expanding at 2400~km/s since the outburst in
2004, i.e., having an inner radius of $1.5\times10^{16}$~cm at the
time of the SN explosion, and $1.6\times10^{16}$~cm at 55~days.  For
the calculations we adopted the following pre-supernova surface
abundances (by number): He : C : N : O = 0.90 : 0.060 : 0.003 : 0.037.
These correspond to a WN/C transition object with an initial mass of
30~M$_{\odot}$ (assuming solar metallicity) (Eldridge \& Vink, 2006).
We note that for a lower metallicity the initial mass of such a star
would be larger while the pre-supernova surface abundances would be
roughly the same.  For example, at an SMC metallicity the initial mass
would be $\sim$50 M$_{\odot}$. In addition, we assumed that hydrogen
was practically absent (at the level of 10$^{-4}$ times the He
abundance) and included Fe at $10^{-3}$ times the He abundance (by
number).\\

An X-ray luminosity of $\sim$4 $\times$ 10$^{39}$ erg~s$^{-1}$ was
observed for SN 2006jc (Immler et al. 2008) at $\sim$100-130 days. We
found that a comparable X-ray luminosity can be produced either using
an ejecta mass of $\sim$10 M$_{\odot}$ and a shell density of $\sim$3
$\times$ 10$^{8}$~cm$^{-3}$ (henceforth Model 1), or an ejecta mass of
$\sim$8 M$_{\odot}$ and a shell density of $\sim$5 $\times$
10$^{8}$~cm$^{-3}$ (Model 2). In both cases $n = 12$. For lower $n$
and/or ejecta mass, the X-ray luminosity becomes too high at
$\sim$100-130 days.  A slightly lower ejecta mass of $\sim$5 M$_{\odot}$
was obtained by Pastorello et al. (2008a) and Tominaga et al. (2008)
via BLC modelling.  Our calculations show that in both Models~1 and 2,
by 55~days a CDS is formed behind the forward shock.  The mass of the
CDS increases as the shock moves through the shell. By 120~days it is
$\sim$0.18 M$_{\odot}$ in Model 1 and $\sim$0.40 M$_{\odot}$ in Model
2.  The total mass of the swept-up CSM in the two models at 120~days is
$\sim$0.58 M$_{\odot}$ and $\sim$0.53 M$_{\odot}$, respectively.  As
the shock moved through the shell, dragging along with it the shocked
gas, the velocity of the CDS increased to $\sim$3000~km/s by 120~days
in both Models 1 and 2. The shock temperatures in the two models are
$\sim 0.9 \times 10^7$~K and $\sim 1.2 \times 10^7$~K, respectively. 
By 120~days, ejecta at $\sim 1.7 \times
10^4$~km/s had reached the shell, while the outer edge of the CSM
shell is at $\sim 1.8 \times 10^{16}$~cm.  By this epoch, the shock
was close to the outer limit of the CSM shell. At this stage it is
likely that the CDS would be rather quickly accelerated by the ejecta,
eventually being fragmented and dispersed. This provides a plausible
explanation for the sudden drop in X-ray luminosity after 120~days
(see Immler et al. 2008).
By 120~days the mass of the CDS reached $\sim$0.2 - 0.4 M$_{\odot}$ in
the two models discussed, corresponding to a cool carbon mass of
$\sim$0.01 - 0.02 M$_{\odot}$ available to form new amorphous carbon
dust within the CDS.\\

We then proceeded to investigate the possible IR emission from the CDS
dust. How would such dust be heated?  Between 55~days and 231~days the
IR energy released by the dust grains was $\sim2\times10^{48}$~erg.
The heat capacity of the grains is small and so the release of latent
heat during the condensation plus the subsequent cooling would yield
negligible IR emission. The ambient CDS gas would also contain
insufficient thermal energy to provide the necessary heating of the
grains.  In contrast, there is more than enough energy in the shock
itself to power the grain IR emission. Some of this energy might be
coupled to the grains via the X-ray emission from the shock. However,
the X-ray luminosity (Immler et al. 2008) is typically only $\sim$1\%
of the NIR luminosity. Thus, it appears that the early-time
UV/optical luminosity of the ejecta is the only plausible means of
maintaining the NIR luminosity of the CDS grains. By day~230, the CDS
radius was at least $2.0\times10^{16}$~cm, i.e. the dust shell
expanded from 6.0 to at least 7.7~light days radius during the NIR
observations.  Even this minimum size of the shell is sufficient for
light travel time to affect the results. Therefore, we estimated the
IR emission from the CDS dust using IR~echo models similar to those
described above.  We note that in the past such IR~echo models have
only been used for pre-existing dust in the CSM and SN~2006jc thus
provides the first case of an IR echo from newly formed dust within
the CSM around a SN.\\

The thickness of the compressed CSM where the dust is formed is much
less than the whole CSM shell thickness. Following calculations by
Chevalier, Blondin \& Emmering (1992) we adopted 1\% of the shell
radius as the thickness of the dust-forming CDS.  Prior to day~55 i.e.
before there was significant shock interaction with the shell, the
dust shell expansion velocity was assumed to be 2400~km/s,
corresponding to a radius of $1.6\times10^{16}$~cm at day~55.  Between
55~days and 120~days the CDS velocity was accelerated uniformly,
reaching a velocity of 3000~km/s.  Thereafter, the CDS was assumed to
coast at 3000~km/s.  However, as indicated above, the exact velocity
and location of the CDS after the shock had passed through the CSM
shell (at $\sim$120~days) is uncertain.  There is likely to be a phase
of acceleration and fragmentation, and so post-120~day radii are
probably lower limits.\\

The dust material was assumed to be amorphous carbon, having an
emissivity law $\lambda^{-1.15}$.  Given that the dust had recently
formed, it was assumed that the grain size would be small (e.g. Nozawa
et al. 2008).  A grain radius of 0.005~$\mu$m was adopted.  However
for grain radii less than about $\lambda/2\pi$ i.e. $<\sim$0.1~$\mu$m
in this situation, the results are quite insensitive to the actual
value chosen. The grain number density growth was represented by \\

$n(t)=n_0(1-exp(-(t-t_0)/t_d))$  \\

\noindent
where $t$ is time, $t_0$ is the time at which dust condensation began,
$t_d$ is the characteristic grain growth timescale, and n$_{0}$ is the
dust number density scaling factor. All times are as viewed from the
supernova. Thus, owing to the light travel time differences, from the
Earth the grain condensation is seen to commence during the epochs
($t_0-6.0$) to ($t_0+6.0$) days.  Within the thin shell the dust
number density was assumed to be uniform. The grains were assumed to
appear instantaneously at their final size. No attempt was made to
simulate the growth of individual grains, although this was probably
fast once conditions were right (cf. a grain growth timescale in ejecta
of a few days: Todini \& Ferrara, 2001, Nozawa et al., 2003).\\

The source of energy for the echo is assumed to be the UV/optical
radiation from the supernova ejecta. For epochs beyond 22~days, we
used a parameterised description of the blackbody-fit bolometric light
curve of Pastorello et al. (2008a).  The SN was unobserved during the
0--22~day period, but this is of no consequence here since this part
of the SN emission had travelled well beyond the shell before grain
condensation commenced.  The parameterised BLC was scaled upwards by
$15\%$. This was to allow for the fact that by about 80~days the IR
luminosity from the shell dominated the BLC.  Consequently the
observed BLC was delayed in the observer's time frame by about 6 days
on average, compared with the SN frame. Thus, the 15\% enhancement has
the effect of moving the light curve time axis back by about 6
days. The SN itself was represented as a point source. Previously
(Table~2) we found that even as early as 55~days, the hot component
blackbody radius was only about 1$\%$ of the shell radius. Thus, it is
likely that the hot photosphere was small compared with the shell.  \\

We found that the dust shell echo model, as described above, was able
to reproduce the $K$ light curve satisfactorily up to about 180~days.
Moreover, the $H$ light curve was also reasonably reproduced by the
model (see Fig.~9). In addition, and of particular note, is that the
rapidly declining equilibrium temperature for potential grains within
the shell fell to 1900~K during the observation period 49--61~days
(the range here being due to light travel time across the shell) (see
Fig.~9, RH panel). This is about the maximum condensation temperature
of amorphous carbon grains.  Thus the shell-echo scenario yields a
natural explanation for the particular epoch at which the NIR emission
appeared.  Prior to $\sim$50 days, the UV/optical light from the SN
rendered the shell environment too hot for grains to form.  We also
found that the MIR SED at epochs 228 and 430~days, and the $K$-band
493~day flux were underproduced by the shell-echo model. As with the
blackbody study (Sect. 4.1) an additional cool component 
appeared to be present.  While it may be possible to modify the CDS
dust model so as to exhibit a range of dust temperatures (e.g. using a
range of dust grain sizes, emissivities and densities) and so account
for the cool, excess flux component, such an explanation would be
rather {\it ad hoc} and have an unclear physical basis.  Other
possibilities are that the cool excess emission arose from (a) new
dust formed in the ejecta or (b) an IR~echo from pre-existing dust in
a circumstellar wind. \\

\subsection{The source of the cool late-time excess emission}
To examine the ejecta dust hypothesis for the cool excess IR emission,
we added a uniform dust sphere model (see Section~4.3) to the
day~231 shell model. We find that to reproduce the excess MIR flux,
the dust sphere must be expanding with a velocity of at least
9000~km/s. For the 9000~km/s case, $T=590$~K, the dust mass is
$1.0\times10^{-3}$~M$_{\odot}$ and the dust is optically thick, even
in the MIR, i.e. the SED is essentially a blackbody.  Good
reproduction of the cool excess flux can also be obtained with similar
masses of optically thinner dust at similar temperatures, but the
expansion velocity has to be increased, reaching $\sim$20,000~km/s by
the time the dust is optically thin in the MIR (although even at this
velocity it is still optically thick in the optical/NIR region). Given
that refractory elements might exist at velocities up to
$\sim$8000~km/s, it is conceivable that newly-formed optically thick
ejecta dust could be responsible for the excess MIR flux. Such a flux
would be powered by the radioactive decay.  Assuming Model~C of
Pastorello et al. (2008a), and using the formulae reported by Valenti
et al. (2008) (see also Clocchiatti \& Wheeler, 1997; Colgate et
al. 1997) to find the deposited luminosity we find that the optically
thick cool excess model luminosity corresponds to $\times1.08$ of the
deposited energy i.e. given the uncertainties, there is just about
enough energy for the ejecta dust hypothesis for the MIR excess at
this epoch. However, for ejecta to reach the shell by day~230 requires
a velocity of just $\sim$10,000~km/s.  Thus, the optically thick
ejecta dust hypothesis implies that its boundary would be almost
contingent with the inner boundary of the shell, and would totally
block out emission from the ejecta and the back half of the shell.
This raises the difficulty that the CDS dust has a temperature of
830~K at this epoch (see Fig. 9) and yet the enclosed, near-contingent
surface responsible for the cool excess emission would have a
temperature of just 590~K.  A variation on the new-ejecta dust
hypothesis for the cool late-time emission might be that a reverse
shock from the ejecta/He~I interaction produced a second, inner CDS
within the ejecta. However, given the high velocity and low densities
involved, we think that such a shock would not produce a significant
CDS. We conclude that the IR emission from ejecta dust is an unlikely
origin for the cool late-time excess flux.\\

An alternative and arguably more natural explanation for the cool
excess flux component is provided by the fact that the progenitor or
its binary companion showed an LBV-like outburst shortly before the
explosion of SN 2006jc.  Such a progenitor system would also be
expected to have undergone longer-term mass-loss producing an extended
circumstellar wind. Therefore, in addition to the IR~echo from the
CDS, we would also expect to see a more conventional IR~echo from
pre-existing dust in the wind.  Indeed, it would be surprising {\it
not} to see any IR~echo at all from this region.  Sakon et al. (2008)
have already suggested that the MIR excess could be due to an IR~echo
from pre-existing CSM dust.  We therefore added a second echo, from
the progenitor wind, to the model. 

\subsection{The two-echo model}
For the progenitor wind component of the two-echo model, a constant
mass loss rate was assumed so the density varied as $r^{-2}$. The dust
material was assumed to be amorphous carbon with an emissivity law
$\lambda^{-1.15}$. The input luminosity was again assumed to be the
UV/optical radiation from the SN ejecta. However, after the formation
of the shell dust, most of the luminosity reaching the wind dust would
have been in the form of IR radiation, which would have been
inefficiently absorbed by the wind dust. Therefore, to make a
conservative allowance for this, the input luminosity was a
parameterised description of just the $UBVRI$ bolometric light curve of
Pastorello et al. (2008a), As in Sect. 4.4, to allow for the unobserved flux
shortward of the $U$-band, the luminosity was scaled up by a factor of
about $\times1.9$, and the early unobserved portion represented using
a plausible extrapolation of the $UBVRI$ BLC (see Sect. 4.4).  The
characteristic wavelength of the input BLC radiation was assumed to be
0.5 $\mu$m.\\

For the CDS shell component of the two-echo model, the dust number density
scaling factor, plus the dust condensation start time and timescale
were adjusted to obtain a match to the 55--231~day $K$-band light
curve.  For the wind component, the grain radius, dust number density
scaling factor and the inner and outer radii of the CSM were
adjusted to match the MIR excess flux and the $K$-band point on
day~493.  The model results are compared with the data in Figures~9
and 10. The model parameter values are summarised in Table~5.
\begin{figure}
\begin{minipage}{85mm}
\includegraphics[width=85mm, angle=0, clip]
{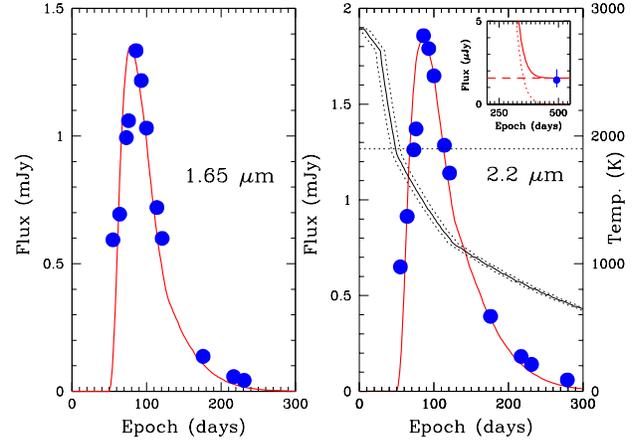}
\caption{Comparison of SN~2006jc observed $H$ and $K$ light curves
with the two-echo model.  During the period shown in the main plots
(0--300~days) the NIR fluxes are dominated by the shell emission.  The
inset in the RH panel shows a magnified plot of the model up to
550~days.  The model is shown as a solid line, while the shell and
wind components are shown as dotted and dashed lines, respectively. It
can be seen that the 493~day $K$~band flux is dominated by the flux
from the CSM wind.  Also shown in the RH panel (plots descending from
top left) is the equilibrium dust temperature.  The three plots (going
left to right) indicate the temperature at the near side (dots),
half-way (solid line), and the far side (dots) of the shell. The
horizontal dotted line indicates a temperature of 1900~K, at which the
condensation of amorphous carbon grains might be expected to begin.
This corresponds to the observation epochs days~49-61.}
\end{minipage}
\end{figure}
\begin{figure}
\begin{minipage}{85mm}
\includegraphics[width=85mm, angle=0, clip]
{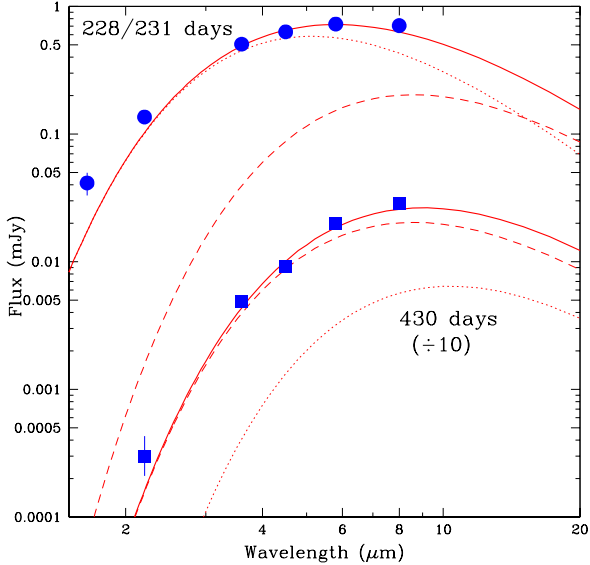}
\caption{Comparison of SN~2006jc observed NIR/MIR SEDs on days 228/31
and 430 with the two-echo model. The model is shown as a solid line,
while the shell and wind components are shown as dotted and dashed
lines, respectively.  For clarity, the day~430 model and data have
been displaced downwards by a factor of 10. The $K$ band point at
430~days was obtained by interpolation between days~279 and 493. Owing
to the uncertainty in this procedure an error of $\pm0.7$ mags. has
been assigned to the interpolated point}
\end{minipage}
\end{figure}
\begin{table*}
\begin{center}
\caption{Parameter values of the two-echo model.}
\begin{tabular}{llllllll}
\hline
Component & $r_{in}$ (cm) & $r_{out}$ & $a$ ($\mu$m) & $t_0$ (d) & $t_d$ (d) & $\tau_{UV-opt}$ & $M_{dust}$ (M$_{\odot}$) \\
\hline
CDS$^*$   & $20.7\times10^{15}$ & $20.9\times10^{15}$ & 0.005 & 50 & 160 & 3.7 & $0.3\times10^{-3}$ \\
Wind      & $750\times10^{15}$ & $3000\times10^{15}$ & 0.13 &  -- & -- & 0.019 & $8.4\times10^{-3}$ \\
\hline
\end{tabular}
\end{center}
\begin{flushleft}
$^*$ The CDS shell radii, optical depth and dust mass are for the epoch 230~days. 
\end{flushleft}
\end{table*}
We note that a fair match to the $H$ band light curve was obtained
without further adjustment of the model.  A range of wind parameter
values and grain sizes allowed the wind component to reproduce the MIR
excess flux on days~228 and 430 plus the day~493 $K$-band point.  The
inner radius, $r_{in}$, could range from $\sim6\times10^{17}$~cm (230
light days) to $\sim1.2\times10^{18}$~cm (460 light days). The
corresponding grain radii were from 0.16~$\mu$m down to 0.05~$\mu$m, while
the corresponding total dust masses ranged from
$8\times10^{-3}$~M${_\odot}$ to $25\times10^{-3}$~M${_\odot}$.  The
model indicates that the evaporation radius would be about
$0.8\times10^{17}$~cm and so it is unlikely that this was the main
cause of the cavity. Instead, we invoke episodic mass loss.  A large
cavity due to episodic mass loss was also inferred in the MIR study of
SN~2002hh (Meikle et al. 2006).  For $r_{in}<0.6\times10^{18}$~cm,
while a match at 430/493~days could be achieved, the earlier NIR flux
was increasingly overproduced.  For $r_{in}>1.2\times10^{18}$~cm the
dust temperature became too low to match the shape of the MIR SED.
Also, above this radius the CSM mass became increasingly implausible
($M_{CSM}>5~$M${_\odot}$).  Satisfactory matches to the data could
also be achieved with a range of wind outer radii, the main constraint
here also being that the wind mass should stay within plausible bounds.
\\

The shell dust mass at 230~days reached $0.3\times10^{-3}$~M$_{\odot}$
i.e. just a few per cent of the cool carbon mass available to form new
amorphous carbon dust within the CDS (see above).  If the dust mass
continued to grow as specified by the exponential factor, it would
asymptotically approach $0.44\times10^{-3}$~M$_{\odot}$.  However, we
have no evidence that the growth continued beyond 230~days.  For the
wind we give results for $r_{in}=7.5\times10^{17}$~cm and
$r_{out}=30\times10^{17}$~cm, yielding a wind dust mass of
$8.4\times10^{-3}$~M$_{\odot}$.  This is close to the lower limit for
the dust mass (see above).  These dust masses correspond to gas masses
of $0.03 \times (0.01/r_{dg})$~M$_{\odot}$ in the shell (at 230~days)
and $0.84 \times (0.01/r_{dg})$~M$_{\odot}$ in the wind, where
$r_{dg}$ is the dust-to-gas mass ratio.  For the pre-existing dust,
the value is comparable to the dust and CSM masses found by Pozzo et
al. (2006) and Meikle et al. (2006) for the Type~IIP SN~2002hh.  We
note also that Morris et al. (1999) used ISO observations extending to
200~$\mu$m to infer 0.15~M$_{\odot}$ of dust in the CSM of Eta
Car. Thus, we consider the dust mass estimate for the SN~2006jc
progenitor CSM as being entirely plausible.  The UV/optical optical
depths are 3.7 and 0.019 for the shell dust and CSM dust
respectively. The rather large optical depth of the shell is in
agreement with the increase in the extinction of A$_{V}$ $\sim$ 3
estimated from the steepening of the optical light curves.  It
confirms that the shell must have been essentially opaque to
UV/optical photons explaining the disappearance of the broad ejecta
lines and the almost complete attenuation of the red wings of the He~I
lines. The low optical depth of the wind is in accord with the low
extinction towards SN~2006jc.  For the grain number density growth
function, $t_0=50$~days and $t_d=160$~days. The quite large value of
$t_d$ is necessary to compensate for the rapid decline of the input SN
light curve. This is, perhaps, a surprisingly long time given that it
probably only took about 120~days for the shock to sweep through the
shell. It may indicate that dust formation continued after the shock
had departed the shell, or that the extent of the dust forming region
was considerably greater than the $1\%$ of the shell radius that was
adopted. \\

In summary, we note that having achieved a satisfactory match to the
$K$ band light curve, the model spontaneously (a) generated an
appropriate condensation temperature in the CDS at the right time, and
(b) reproduced the $H$ band light curve. This was achieved using the
actual supernova bolometric light curve as input, an amorphous carbon
emissivity, a simple grain number density growth scenario, and a
two-component spherically symmetric IR~echo model. The MIR excess and
the 493~day $K$-band point are satisfactorily reproduced by a dusty
progenitor wind.  Given the nature of the progenitor system, this
seems entirely plausible. It seems unlikely that the cool excess IR
emission arose solely from dust formed in the same CDS that produced
the NIR emission. It is also unlikely that the cool excess emission
was produced by newly-formed ejecta dust powered by radioactive decay
or by a reverse shock.  In our view, a combination of IR~echoes from
the CDS dust and from a more extended dusty progenitor wind yield the
most complete and convincing explanation for the IR behaviour of
SN~2006jc.\\

\section{Discussion}
Smith et al. (2008) find that only $0.6\times10^{-5}$~M$_{\odot}$ of dust is
needed to account for the NIR luminosity. At the epoch of their
observations (95~days, our epoch definition) we find a shell dust mass
of $1.1\times10^{-4}$~M$_{\odot}$.  We suspect that some of this factor
of 18 discrepancy is due to the much larger grain radius, 0.3~$\mu$m,
used by Smith et al. compared with our 0.005~$\mu$m.  Inspection of
Draine \& Lee (1984) shows that for amorphous carbon, $Q_\nu/a$ is
about $\times3.5$ larger at $a=0.3\mu$m radius than for $a<0.03\mu$m,
and so the larger grain size will yield the same luminosity with less
than 30\% of the mass. This is due to the increasing contribution of
the magnetic dipole term to the emissivity as $a$ increases above
0.03~$\mu$m.  To our knowledge, grain growth calculations for a
SN-shocked circumstellar shell have not yet been performed.  However,
we note that for grain growth in the SN ejecta environment the grain
size is likely to be less than about 0.05~$\mu$m (e.g. Todini \&
Ferrara, 2001; Nozawa et al. 2008).  It therefore seems possible
that the dust mass of Smith et al. required to account for the NIR
luminosity is an underestimate. In addition, a grain size as large as
0.3~$\mu$m might not produce even the modest reddening within the
optical region indicated by the optical light curves (Section
4.2).\\

Smith et al. (2008) also suggest that the total dust mass
produced in the CDS over a time scale of 2 months could be as high as
1\% of a solar mass or more, assuming that only the very hottest dust
was detectable in the NIR.  Our modelling (see Sect. 4.7) indicates
that at 150 days, the mass of newly-formed dust in the shell required
to account for the NIR flux is about $2\times10^{-4}$ M$_{\odot}$ and
the UV/optical depth is already around 2.5.  As we point out in
Sect. 4.7, this is consistent with the observed optical SN light curve
behaviour. However, having $\sim$0.01 M$_{\odot}$ of dust within the
CDS would result in an enormous optical depth and the dust would
totally block out the optical SN light.  Furthermore, the hot dust
would lie on the {\it inside} of the CDS (heated by the SN luminosity)
and so the NIR light would also be absorbed by the dust shell.
However, as our observations show the SN remained detectable until
$\sim$180 days and $\sim$490 days at optical and NIR wavelengths,
respectively.  Therefore, we find such an enormous dust mass
unlikely.\\

Sakon et al. (2008), Tominaga et al. (2008) and Nozawa et al. (2008)
invoke dust condensation in the ejecta to account for the NIR
luminosity plus part of the MIR luminosity.  Sakon et al. use a
uniform, optically-thin (in the NIR) dust sphere to estimate the dust
mass and temperature at 220~days.  While they do not indicate the size
of this sphere, we find that, given their temperature of 800~K, the
radius must be about $4.7\times10^{16}$cm. To reach
this radius at 220~days would require a velocity of about
25,000~km/s. Even assuming an exceptionally massive progenitor for
SN~2006jc, it seems rather unlikely that refractory elements will
exist at this velocity.  The very early appearance of dust might also
be taken as an argument against ejecta dust formation, but modelling
by Nozawa et al. (2008) suggests that about a solar mass of dust could
have formed in the ejecta of SN~2006jc and that dust formation could
have begun as early as 50~days. However, as they concede, it is
difficult to see how effects such as clumping or destruction by high
energy photons/electrons could account for the much smaller (over 3
orders of magnitude) observed mass.\\

Nozawa et al. (2008) also argue that the LBV-ejected shell density
would be too low for grains to nucleate. They base this on (a) the
X-ray observations of Immler et al. (2008) and (b) the hydrodynamic
calculations of Tominaga et al. (2008).  On the other hand, and as
already pointed out in Pozzo et al. (2004) for SN~1998S and in
Smith et al. (2008) for SN 2006jc, the physics of the
CDS dust formation invoked here is reminiscent of the radiative shock
of colliding winds of Wolf-Rayet stars which is known to be a
dust-forming site (Usov 1991).  Moreover, we have shown that a simple
IR~echo model involving new dust in a CDS and old dust in the
progenitor wind can naturally account for the flux, the NIR/MIR SED
and the evolution of the IR emission from SN~2006jc. As shown above,
using an independently determined bolometric light curve and shell
radius, the shape and rate of decline of the NIR flux are produced
naturally within the IR~echo scenario. In addition, the epoch
predicted by the model at which dust condensation in the shell first
becomes possible coincides with the epoch at which the NIR excess
first appeared. These results give us confidence in the reality of the
IR~echo origin of the IR flux.  Within their ejecta condensation
model, Nozawa et al. find carbon dust condensation beginning at
40--60~days, but it is not clear how dependent this might be on the
details of the sequence of hydrodynamic, nucleosynthesis and condensation
model calculations upon which this result is based. \\

Smith et al. (2008) have argued against a NIR~echo from pre-existing
CSM dust on the basis of the grain temperature and its decay
timescale.  However, we find that the temperature and decay
timescale could indeed be accounted for within a single, pre-existing
dust IR-echo scenario (see Fig.~7).  In our view, the key objections
to a pre-existing dust IR~echo as the sole origin of the SN~2006jc IR
behaviour are (a) the insufficient luminosity of the SN BLC, (b) the
delayed rise in the NIR light curves, and (c) the late-time MIR flux.
Smith et al. have also argued that a pre-existing dust IR~echo
scenario would not account for the attenuation of the He~I line
profiles. We agree that this is true for an echo from an extended
progenitor wind. We also agree that the CDS dust would be
able to produce the observed He~I profile evolution. As we show, an
IR~echo from this CDS dust can account for the bulk of the near-IR
emission from SN~2006jc. The dependence of the grain number density
growth on its equilibrium temperature within the IR~echo model also
explains why the IR excess did not appear earlier. (The time it took
for the SN shock to reach the shell probably also constrained the
epoch at which dust was able to condense.) \\

\section{Conclusions}
The discovery of the IR excess in SN 2006jc has provided us with, for
the first time, an opportunity to study in detail this phenomenon in
an H-deprived SN.  We have shown that the interaction of the ejecta
outward shock with a dense shell of material ejected by the progenitor
in an LBV-like outburst about two years prior to the SN explosion was
able to produce a CDS behind the forward shock by 55 days from the
explosion.  The intensity, SED and evolution of the IR flux together
with other evidence leads us to propose that this emission was due to
IR~echoes. The bulk of the NIR flux came from newly-formed CDS dust
while a substantial and growing fraction of the MIR flux came from
pre-existing dust in the progenitor wind, probably lying beyond
$6\times10^{17}$~cm.  The CDS amorphous carbon dust mass was
$0.3\times10^{-3}$~M$_{\odot}$ which is just a few percent of the cool
carbon mass of $\sim$0.01--0.02 M$_{\odot}$ at 120~days. This model 
explains the observed NIR evolution, as well as providing enough extinction 
($A_V\sim3$) to account for the fast decline in the optical light curves, the
attenuation of the red wings of the CSM He~I emission lines, and also
the disappearance of the ejecta lines.  The mass of pre-existing dust in the 
wind was at least $\sim8\times10^{-3}$~M$_{\odot}$.  Given the
$\sim6\times10^{17}$~cm lower limit for $r_{in}$ derived for the
pre-existing CSM dust, and assuming a typical Wolf Rayet star wind
velocity of 1000~km/s, we can infer that the episodic mass-loss phase
ceased at least $\sim$200 years before the pre-SN outburst and the
explosion of SN~2006jc.  For the wind model results presented above
(Table~5) the mass loss would have taken place during the period
240--950 years before the SN explosion, implying a mass-loss rate of
$1.2\times10^{-3}(0.01/r_{dg})$~M$_{\odot}$/year.  This is rather high
for a WR star (see e.g. Eldridge et al. 2006), adding weight to the
proposition that SN 2006jc had an unusual progenitor.\\

The IR behaviour of SN~2006jc can be explained as a combination of
IR~echoes from two manifestations of stellar mass loss and this work
provides two main conclusions.  Firstly, it adds to the growing
evidence that mass-loss from the {\it progenitors} of core-collapse
supernovae could be a major source of dust in the universe.  Secondly,
we have witnessed dust formation in yet another type of core-collapse
supernova. Furthermore, while dust condensation within the CDS formed 
behind the ejecta inward shock (in mainly ejecta material) has been 
proposed before for one event (SN 1998S), SN 2006jc is the first case with
evidence for dust condensation in the CDS behind the ejecta outward
shock in the circumstellar material.  Finally, we note that two other
events (SNe 1999cq and 2002ao) of the same Type Ibn class have also
shown steepening optical light curves, similar to those of SN 2006jc.
This suggests that CDS dust formation might well be a common
characteristic in other events of this SN type (see also Smith et
al. 2008 and Pastorello et al. 2008a).  However, once again, we have
seen no direct evidence that the explosion of a supernova produces
anything other than a very modest amount of dust. \\

\section*{Acknowledgements}
We thank an anonymous referee for useful comments and S. Valenti, L. 
Zampieri and D. Watson for helpful discussions.
We also thank the Cambridge Astronomical Survey Unit (CASU) for access to
the reduced UKIRT WFCAM images.  This work is based on observations made
with the United Kingdom Infrared Telescope (UKIRT), the Spitzer Space
Telescope, and the Gemini Observatory.  UKIRT is operated by the Joint
Astronomy Centre on behalf of the Science and Technology Facilities
Council of the U.K.  The Spitzer Space Telescope is operated by the
Jet Propulsion Laboratory, California Institute of Technology, under a
contract with NASA.  The Gemini Observatory is operated by the
Association of Universities for Research in Astronomy, Inc., under a
cooperative agreement with the NSF on behalf of the Gemini
partnership: the National Science Foundation (United States), the
Science and Technology Facilities Council (United Kingdom), the
National Research Council (Canada), CONICYT (Chile), the Australian
Research Council (Australia), CNPq (Brazil) and SECYT (Argentina).
Financial support for this work was provided by NASA through awards
(30292, 40619) issued by JPL /Caltech.  This work, conducted as part
of the award 'Understanding the lives of massive stars from birth to
supernovae' made under the European Heads of Research Councils and
European Science Foundation EURYI Awards scheme, was supported by the
Participating Organisations of EURYI and the EC Sixth Framework
Programme. SM acknowledges financial support from the Academy of
Finland (project: 8120503).

\end{document}